\documentclass[ijoc,nonblindrev]{informs3} % current default for manuscript submission

\OneAndAHalfSpacedXII % current default line spacing
\usepackage{algorithm}
\usepackage[noend]{algorithmic}

\usepackage[dvipsnames]{xcolor}
\usepackage{tikz}
\usetikzlibrary{fit,positioning,shapes.geometric,shapes.misc,arrows.meta,decorations.markings,external}
\usepackage{hyperref}
\usepackage{cleveref}

\Crefname{ALC@unique}{Line}{Lines}
\newcounter{myalg}
\AtBeginEnvironment{algorithmic}{\refstepcounter{myalg}}
\makeatletter
\@addtoreset{ALC@unique}{myalg}
\makeatother
\usepackage{color}

\newenvironment{myproof}
    {\proof{Proof.}}
    {\Halmos\endproof}

\usepackage{framed}
\definecolor{shadecolor}{RGB}{230,230,230}
\newenvironment{important}
	{\vspace{-5pt}\begin{snugshade}}
	{\end{snugshade}\vspace{-5pt}}

\newcommand{\lst}[1]{\left(#1\right)}
\newcommand{\vct}[1]{\left\langle#1\right\rangle}
\newcommand{\set}[1]{\left\{#1\right\}}
\newcommand{\fromto}[2]{#1,\dots,#2}
\newcommand{\dd }[1]{\mathcal{#1}}
\newcommand{\arc}[3][]{(#2\xrightarrow{#1}#3)}
\newcommand{\pth}[2]{#1\rightsquigarrow#2}

\newcommand{\lb}[1]{\underline{#1}}
\newcommand{\ub}[1]{\overline{#1}}
\newcommand{\opt}[1]{#1^*}
\newcommand{\rlx}[1]{\ub{\dd{#1}}}
\newcommand{\rst}[1]{\lb{\dd{#1}}}
\newcommand{\apx}[1]{\dd{#1}}
\newcommand{\rub}[1]{\ub{v}_{rub}(\sigma(#1))}
\newcommand{\locb}[1]{\ub{v}_{locb}(#1 \mid \rlx{B})}

\newcommand{\pb}[1]{\mathcal{#1}}

\usepackage{natbib}
 \bibpunct[, ]{(}{)}{,}{a}{}{,}%
 \def\bibfont{\small}%
\TheoremsNumberedThrough     % Preferred (Theorem 1, Lemma 1, Theorem 2)
\EquationsNumberedThrough    % Default: (1), (2), ...
\begin{document}
%%%%%%%%%%%%%%%%

% Outcomment only when entries are known. Otherwise leave as is and 
%   default values will be used.
%\setcounter{page}{1}
%\VOLUME{00}%
%\NO{0}%
%\MONTH{Xxxxx}% (month or a similar seasonal id)
%\YEAR{0000}% e.g., 2005
%\FIRSTPAGE{000}%
%\LASTPAGE{000}%
%\SHORTYEAR{00}% shortened year (two-digit)
%\ISSUE{0000} %
%\LONGFIRSTPAGE{0001} %
%\DOI{10.1287/xxxx.0000.0000}%

% Author's names for the running heads
% Sample depending on the number of authors;
% \RUNAUTHOR{Jones}
% \RUNAUTHOR{Jones and Wilson}
% \RUNAUTHOR{Jones, Miller, and Wilson}
% \RUNAUTHOR{Jones et al.} % for four or more authors
% Enter authors following the given pattern:
\RUNAUTHOR{Copp\'{e}, Gillard and Schaus}

% Title or shortened title suitable for running heads. Sample:
% \RUNTITLE{Bundling Information Goods of Decreasing Value}
% Enter the (shortened) title:
\RUNTITLE{DD-Based Branch-and-Bound with Caching for Dominance and Suboptimality Detection}

% Full title. Sample:
% \TITLE{Bundling Information Goods of Decreasing Value}
% Enter the full title:
\TITLE{Decision Diagram-Based Branch-and-Bound with Caching for Dominance and Suboptimality Detection}
%Towards Disjoint Branching in\\DD-Based Branch-and-Bound

% Block of authors and their affiliations starts here:
% NOTE: Authors with same affiliation, if the order of authors allows, 
%   should be entered in ONE field, separated by a comma. 
%   \EMAIL field can be repeated if more than one author
\ARTICLEAUTHORS{%
\AUTHOR{Vianney Copp\'{e}, Xavier Gillard, Pierre Schaus}
\AFF{UCLouvain, Louvain-la-Neuve, Belgium\\ \EMAIL{\{vianney.coppe,xavier.gillard,pierre.schaus\}@uclouvain.be}}
} % end of the block

\ABSTRACT{%
The branch-and-bound algorithm based on decision diagrams introduced by Bergman et al. in 2016 is a framework for solving discrete optimization problems with a dynamic programming formulation.
It works by compiling a series of bounded-width decision diagrams that can provide lower and upper bounds for any given subproblem.
Eventually, every part of the search space will be either explored or pruned by the algorithm, thus proving optimality.
This paper presents new ingredients to speed up the search by exploiting the structure of dynamic programming models.
The key idea is to prevent the repeated expansion of nodes corresponding to the same dynamic programming states by querying expansion thresholds cached throughout the search.
These thresholds are based on dominance relations between partial solutions previously found and on the pruning inequalities of the filtering techniques introduced by Gillard et al. in 2021.
Computational experiments show that the pruning brought by this caching mechanism allows significantly reducing the number of nodes expanded by the algorithm.
This results in more benchmark instances of difficult optimization problems being solved in less time while using narrower decision diagrams.
}%

% Sample 
%\KEYWORDS{deterministic inventory theory; infinite linear programming duality; 
%  existence of optimal policies; semi-Markov decision process; cyclic schedule}

% Fill in data. If unknown, outcomment the field

\maketitle
%%%%%%%%%%%%%%%%%%%%%%%%%%%%%%%%%%%%%%%%%%%%%%%%%%%%%%%%%%%%%%%%%%%%%%

% Text of your paper here

\section{Introduction}

Many NP-hard combinatorial optimization problems can be modeled and solved with \textit{dynamic programming} (DP).
Although this resolution technique is appealing, the process of memorizing all solved subproblems rapidly becomes infeasible in terms of memory, as the required memory size can grow exponentially with the input size, rendering the computation intractable.
Discrete optimization with \textit{decision diagrams} (DDs) is a recent framework for addressing the memory issue for solving constraint optimization problems using their DP formulation.
Apart from offering new modeling perspectives, this technique can exploit the compactness of DP models within an adapted \textit{branch-and-bound} (B\&B) algorithm introduced by \cite{bergman2016discrete}.
In addition to conducting the search within the DP state space, the strength of DD-based B\&B lies in its dedicated approach to deriving lower and upper bounds.
Both are obtained by compiling bounded-width DDs, respectively called \textit{restricted} and \textit{relaxed} DDs -- assuming a maximization problem.
Those approximate DDs each have their own strategy for limiting their size, which works either by removing or merging excess nodes.

The DDs used today in the field of discrete optimization originate from compact encodings of Boolean functions, first known as binary decision programs \citep{lee1959representation} and later as \textit{binary decision diagrams} (BDDs) \citep{akers1978binary,bryant1986graph}.
\textit{Multi-valued decision diagrams} (MDDs) were then suggested by \cite{kam1990multi} as an extension of BDDs to variables and functions taking values from discrete sets.
These different variants of DDs were successfully applied in different domains such as formal verification \citep{hu1995techniques}, model checking \citep{clarke1994model}, computer-aided design \citep{minato1995binary} and optimization \citep{lai1994evbdd,Hachtel1997,becker2005bdds,hadzic2006,hadzic2007}.
More recently, \cite{andersen2007constraint} introduced relaxed DDs that act as a constraint store for constraint programming solvers.
These DDs can represent a superset of the feasible variable assignments with a bounded width that balances the computational costs and the filtering strength.
Relaxed DDs were then adapted by \cite{bergman2014optimization} as a mean of deriving upper bounds for discrete optimization problems that can be modeled with DP \citep{hooker2013decision}.
They were followed closely by their lower bounding counterparts called restricted DDs \citep{bergman2014bdd}, which represent a subset of the solution space.

Several improvements to the B\&B algorithm based on these two ingredients have been suggested since its introduction: \cite{gillard2021improving} proposed additional bounding procedures to enhance the pruning of the B\&B and thus speed up the search.
Another promising research direction is to discover variable orderings that yield approximate DDs with better bounds.
\cite{cappart2022improving} designed a reinforcement learning approach to perform this task while \cite{karahalios2022variable} proposed different portfolio mechanisms to dynamically select the best ordering among a predefined set of alternatives.
For a complete overview of the latest theoretical contributions and applications of DDs in the field of discrete optimization, we refer the reader to the survey by \cite{castro2022decision}.

The contributions of this paper stem from a simple observation: as opposed to classical B\&B for \textit{mixed-integer programming} (MIP), the DD-based B\&B does not split the search space into disjoint parts.
The reason is that in this framework, DDs are based on a DP formulation of the discrete optimization problem at hand, and such kind of model typically contains many overlapping subproblems.
As a result, the B\&B may explore some subproblems multiple times.
In addition, the approximate DDs compiled during the algorithm can repeat a lot of the work done previously because they cover overlapping parts of the search space.
This inability to build on previous computational efforts is unfortunate, for that is the very strength of the DP paradigm.
A first attempt to address this problem was made in \citep{coppe_et_al:LIPIcs.CP.2022.14} by changing the ordering of the nodes in the B\&B.
By employing a breadth-first ordering, it is guaranteed that at most one B\&B node corresponding to each given subproblem will be explored.
However, this approach sacrifices the benefits of best-first search -- always exploring the most promising node to try to improve the incumbent solution and tighten the bounds at the same time -- and leaves the issue of overlapping approximate DDs unsolved.

In this paper, the same problem is tackled while allowing a best-first ordering of the B\&B nodes.
In the same fashion that a closed list prevents the re-expansion of nodes in shortest-path algorithms, we propose to maintain a \textit{Cache} that stores an \textit{expansion threshold} for each DP state reached by an exact node of any relaxed DD compiled during the B\&B.
These expansion thresholds combine \textit{dominance} and \textit{pruning thresholds} that exploit the information contained in relaxed DDs to the fullest extent.
The former detect dominance relations between partial solutions while the latter extrapolate the pruning decisions related to nodes filtered by the \textit{rough upper bounds} and \textit{local bounds} introduced by \cite{gillard2021improving}.
By consulting the expansion thresholds stored in the \textit{Cache}, many partial solutions are guaranteed to be either dominated by other partial solutions previously found, or suboptimal with respect to bounds already obtained, and can therefore be pruned during the compilation of subsequent approximate DDs.

In the same line of research, \cite{rudich_et_al:LIPIcs.CP.2022.35} suggested inserting open relaxed DDs directly in the B\&B queue instead of open nodes, and described a \textit{peeling} operation that splits a relaxed DD into two parts: the first containing all the paths traversing a chosen node and the second containing the rest.
The peeling operation permits two things when used as a replacement for classical branching procedures: it allows warm-starting the compilation of subsequent relaxed DDs, and it strengthens the bounds of the relaxed DD on which the peeling was conducted.
This mechanism helps in reducing the overlap between the approximate DDs compiled during the search but does not fully address the issue.
These improvements could thus be combined with the techniques hereby introduced.

The paper begins with a summary of DD-based B\&B in \Cref{section:preliminaries}, starting from the DD representation of a discrete optimization problem to the B\&B algorithm and its latest improvements.
This general introduction is followed by a discussion of the caveats that this paper tries to address in \Cref{section:caveats}.
Next, \Cref{section:cache} presents the expansion thresholds through its two components -- the dominance and pruning thresholds -- as well as their computation and integration into both the B\&B algorithm and the approximate DD compilation procedure.
\Cref{section:limitations} then discusses the limitations of the techniques introduced in the paper.
Finally, experimental results on three different discrete optimization problems are reported and discussed in \Cref{section:experiments} before concluding.

\section{Preliminaries}
\label{section:preliminaries}

This section provides an overview of DD-based discrete optimization.
It starts with a reminder on how discrete optimization problems are modeled and solved with DP.
Next, the translation of a DP formulation to DDs is explained.
In particular, \textit{restricted} and \textit{relaxed} DDs are defined as well as the top-down compilation algorithm used to obtain them.
The section ends with a description of the DD-based B\&B algorithm exploiting the lower and upper bounds derived from restricted and relaxed DDs, respectively.
Throughout the section, a simple \textit{Bounded Knapsack Problem} (BKP) scenario is developed to help the reader visualize the concepts presented.

\subsection{Dynamic Programming Formulation}
\label{section:dp}

A discrete optimization problem $\pb{P}$ is defined by $\max \set{ f(x) \mid x \in Sol(\pb{P}) = D \cap C }$, where $x = \lst{\fromto{x_0}{x_{n-1}}}$ is a vector of variables taking values in their respective domains $x_j \in D_j$ with $D = D_0 \times \dots \times D_{n-1}$, $C$ is a set modeling all the constraints that solutions need to satisfy and $f$ is the objective function to be maximized.
To formulate $\pb{P}$ with DP, the following elements are needed:
\begin{itemize}
    \item the \textit{control variables} $x_j \in D_j$ with $j \in \set{\fromto{0}{n-1}}$.
    \item a \textit{state space} $S$ partitioned into $n+1$ sets $\fromto{S_0}{S_n}$ that correspond to distinct stages of the DP model.
    In particular, $S_j$ contains all states having $j$ variables assigned.
    Several special states are also defined: the \textit{root} $\hat{r}$, the \textit{terminal} $\hat{t}$ and the \textit{infeasible} state $\hat{0}$.
    \item the \textit{transition functions} $t_j : S_j \times D_j\rightarrow S_{j+1}$ for $j=\fromto{0}{n-1}$ that specify how states of consecutive stages are connected.
    \item the \textit{transition value functions} $h_j : S_j \times D_j\rightarrow \mathbb{R}$ for $j=\fromto{0}{n-1}$ that compute the contribution of a transition to the objective.
    \item a \textit{root value} $v_r$ to account for constant terms in the objective.
\end{itemize}

If those elements can be defined for $\pb{P}$, then finding the optimal solution is equivalent to solving:
\begin{equation}
\begin{split}
    \mbox{maximize~} & f(x) = v_r + \sum_{j=0}^{n-1} h_j(s^j,x_j) \\
    \mbox{subject to~}& s^{j+1} = t_j(s^j,x_j)\mbox{, for all }j=\fromto{0}{n-1}\mbox{, with }x_j\in D_j\\
        & s^j \in S_j, j=\fromto{0}{n} \mbox{ and }x \in C.
\end{split}
\end{equation}

\begin{example}
\label{ex:knapsack-dp}
Given a set of items $N=\set{\fromto{0}{n-1}}$ with weights $W = \lst{\fromto{w_0}{w_{n-1}}}$, values $V = \lst{\fromto{v_0}{v_{n-1}}}$ and quantities $Q = \lst{\fromto{q_0}{q_{n-1}}}$, the goal of the BKP is to choose the number of copies of each item to include in the knapsack so that the total value is maximized, and the total weight is kept under a given capacity $C$.
In the DP formulation of this problem, we associate a control variable $x_j \in \set{\fromto{0}{q_j}}$ to each item $j$ that decides the number of copies of it to include in the knapsack.
States are uniquely identified by the remaining capacity of the knapsack.
The state space is thus $S=\set{\fromto{0}{C}}$, with in particular the root state $\hat{r} = C$ and its value $v_r = 0$.
The transition functions are given by:
\begin{equation*}
t_j(s^j,x_j) = \left\{ \begin{array}{lcl}
    s^j - x_jw_j, && \text{if $s^j \ge x_jw_j$,} \\
    \hat{0}, && \text{otherwise.} 
\end{array} \right.
\end{equation*}
The weight of item $j$ is subtracted from the capacity as many times as the number of copies selected.
If the remaining capacity does not allow including $x_j$ copies of item $j$, the transition is redirected to the infeasible state $\hat{0}$.
Similarly, the value of item $j$ is added to the objective for each copy with the transition value functions $h_j(s^j,x_j) = x_jv_j$.
\end{example}

\subsection{Decision Diagrams}
\label{section:dds}

\begin{figure}[t]
    \noindent
    \begin{minipage}{.3\textwidth}
        \newcommand{\tablespacing}{2.85cm}
        \centering
        \vspace{5.45cm}
        \begin{tabular}{c|c|c}
        \multicolumn{1}{c}{$V$} & \multicolumn{1}{c}{$W$} & \multicolumn{1}{c}{$Q$} \\ \hline
        2 & 4 & 1 \\
        3 & 6 & 1 \\
        6 & 4 & 2 \\
        6 & 2 & 2 \\
        1 & 5 & 1 \\ \hline
        \multicolumn{3}{c}{$C=15$}                           
        \end{tabular}\\\vspace{0.25cm}
        (a) BKP instance
    \end{minipage}%
    \begin{minipage}{.7\textwidth}
        \centering
        \input{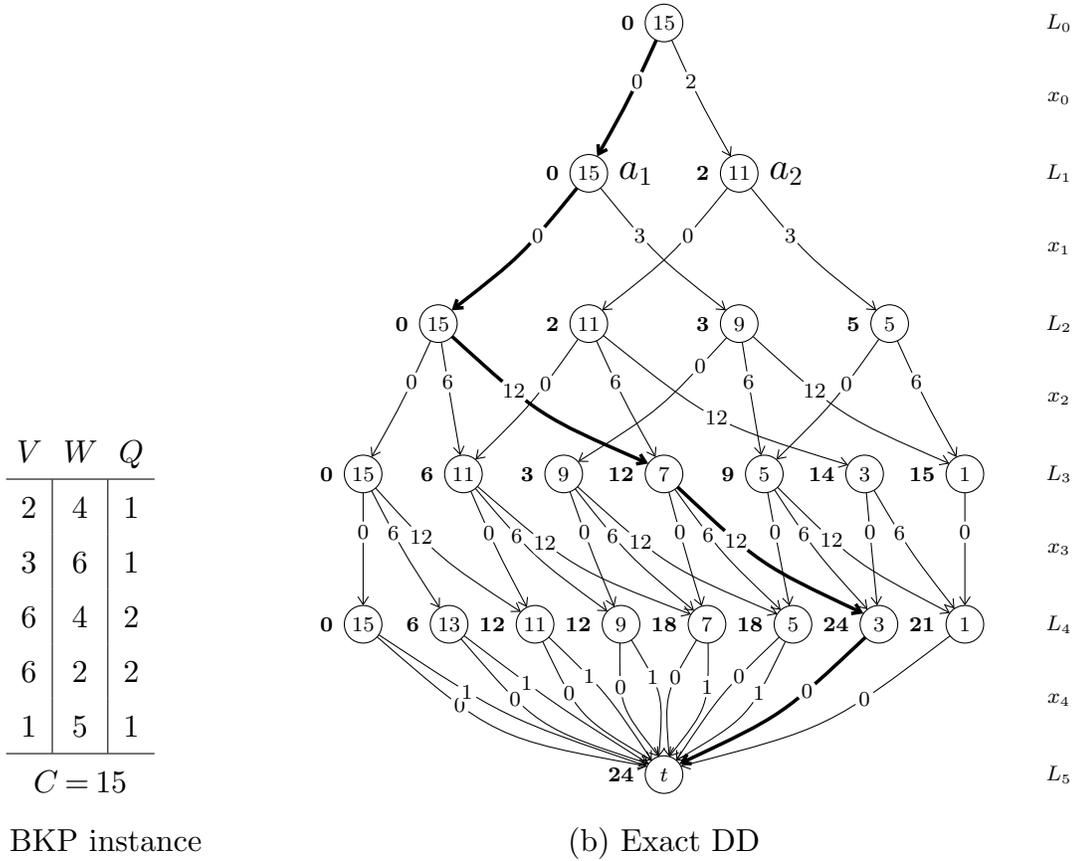}
    \end{minipage}
    \caption{(a) A BKP instance and (b) the corresponding exact DD.
        The value inside each node corresponds to its state -- the remaining capacity -- and the annotation on the left gives the value of the longest path that reaches it.
        For clarity, only arc values are present.
        The longest path is highlighted in bold.}
    \label{fig:exact-dd}
\end{figure}

In the context of discrete optimization, a DD is a graphical representation of the set of solutions to a given problem $\pb{P}$.
DDs are well suited to encode and manipulate compact formulations such as DP models, due to their ability to preserve the uniqueness of overlapping subproblems.
In mathematical terms, a decision diagram $\dd{B}=\lst{U,A,\sigma,l,v}$ is a layered directed acyclic graph consisting of a set of nodes $U$ that are connected by a set of arcs $A$.
Each node is mapped to a DP state by the function $\sigma$.
The set of nodes $U$ can be partitioned into \textit{layers} $\fromto{L_0}{L_n}$ corresponding to the successive stages of the DP model, each containing one node for each \textit{distinct} state of the corresponding stage.
Therefore, arcs $a = \arc[d]{u_j}{u_{j+1}}$ connect nodes of consecutive layers $u_j \in L_j,u_{j+1} \in L_{j+1}$ and represent the transition between states $\sigma(u_j)$ and $\sigma(u_{j+1})$.
The \textit{label} $l(a) = d$ of an arc encodes the decision that assigns the value $d \in D_j$ to variable $x_j$.
The value $v(a)$ of the arc stores the transition value.
Both the first and last layer -- $L_0$ and $L_n$ -- contain a single node, respectively the \textit{root} $r$ and the \textit{terminal} node  $t$.
Each $\pth{r}{t}$ path $p=\lst{\fromto{a_0}{a_{n-1}}}$ that traverses the DD from top to bottom through arcs $\fromto{a_0}{a_{n-1}}$ represents a solution $x(p) = \lst{\fromto{l(a_0)}{ l(a_{n-1})}}$ to $\pb{P}$.
Its objective value is given by the accumulation of the arc values along the path: $v(p) = v_r + \sum_{j=0}^{n-1} v(a_j)$.
Finally, the set of all solutions appearing in the DD is defined as $Sol(\dd{B}) = \set{x(p) \mid \exists p : \pth{r}{t}, p \in \dd{B} }$ and $\dd{B}$ is said \textit{exact} if $Sol(\dd{B}) = Sol(\pb{P})$ and $v(p) = f(x(p)), \forall p \in \dd{B}$.
For convenience, all nodes $u$ and paths $p$ appearing in a DD $\dd{B}$ can be accessed respectively with the notation $u \in \dd{B}$ and $p \in \dd{B}$.

\begin{example}
\label{ex:knapsack-exact-dd}
\Cref{fig:exact-dd}(a) defines a small BKP instance with 5 items and a knapsack capacity of 15.
The exact DD obtained by developing the DP model introduced in \Cref{ex:knapsack-dp} for this instance is represented on \Cref{fig:exact-dd}(b).
The bold path is the longest path of the DD and corresponds to the optimal solution $\opt{x} = \lst{0, 0, 2, 2, 0}$ with value $f(\opt{x}) = 24$.
\end{example}

\begin{algorithm}[t]
\begin{algorithmic}[1]
\STATE $i \gets$ index of the layer containing $u_r$
\STATE $L_i \gets \set{u_r}$ \label{init-first-layer}
\FOR{$j=i$ \textbf{to} $n-1$}
    \STATE $pruned \gets \emptyset$
    \begin{important}
    \IF[pruning for the root done in \Cref{alg:b-and-b}]{$j > i$} \label{ignore-first-layer}
        \FORALL{$u \in L_j$} \label{cache-pruning-start}
            \IF{$Cache.contains(\sigma(u))$ \textbf{and} $\opt{v}(u \mid \dd{B}) \le \theta(Cache.get(\sigma(u)))$} \label{forbidden-transition}
                \STATE $pruned \gets pruned \cup \set{u}$ \label{cache-pruning-end}
            \ENDIF
        \ENDFOR
    \ENDIF
    \end{important}
    \STATE $L_j' \gets L_j \setminus pruned$ \label{remove-pruned}
    \IF{$\lvert L_j' \rvert > W$} \label{max-width-exceeded}
        \STATE restrict or relax the layer to get $W$ nodes with \Cref{alg:layer-reduce} \label{restrict-or-relax}
    \ENDIF
    \STATE $L_{j+1} \gets \emptyset$
    \FORALL{$u \in L_j'$} \label{develop-layer-start}
        \IF[rough upper bound pruning]{$\opt{v}(u \mid \dd{B}) + \rub{u} \le \lb{v}$} \label{rub-pruning}
            \STATE \textbf{continue}
        \ENDIF
        \FORALL{$d \in D_j$}
            \STATE create node $u'$ with state $\sigma(u') = t_j(\sigma(u),d)$ or retrieve it from $L_{j+1}$
            \STATE create arc $a=\arc[d]{u}{u'}$ with $v(a) = h_j(\sigma(u),d)$ and $l(a) = d$
            \STATE add $u'$ to $L_{j+1}$ and add $a$ to $A$ \label{layer-add} \label{develop-layer-end}
        \ENDFOR
    \ENDFOR
\ENDFOR
\STATE merge nodes in $L_n$ into terminal node $t$ \label{merge-terminal}
\end{algorithmic}
\caption{Compilation of DD $\dd{B}$ rooted at node $u_r$ with maximum width $W$.}
\label{alg:dd-compilation}
\end{algorithm}

Given the subproblem $\pb{P}\lvert_{u_r}$ that restricts $\pb{P}$ to paths traversing some node $u_r$, \Cref{alg:dd-compilation} details the top-down compilation of a DD rooted at $u_r$, given that $\sigma(u_r)$ is a state at the $i$-th stage of the DP model.
The notations utilized in the algorithms are deliberately less formal to make the explanations easier to understand.
Additionally, the highlighted portions in \Cref{alg:dd-compilation,alg:b-and-b} pertain to the enhancements introduced in \Cref{section:cache}, which will be explained in further detail later.
After initializing the first layer at \cref{init-first-layer} with the root node only, the DD is developed layer by layer by applying every possible transition to each node of the last completed layer at \crefrange{develop-layer-start}{develop-layer-end}.
Once the DD is fully unrolled, \cref{merge-terminal} merges all nodes of the terminal layer into a single terminal node $t$.
At \cref{max-width-exceeded}, the \textit{width} of the current layer is computed and compared against a parameter $W$ called the \textit{maximum width}.
When the layer width exceeds $W$, the algorithm yields approximate DDs by either restricting or relaxing the layer at \cref{restrict-or-relax}.
In order to obtain an exact DD, one can simply set $W=\infty$.
For most combinatorial optimization problems, the size of the corresponding exact DD can grow exponentially, making it intractable to compute.
Restricted and relaxed DDs are both capable of maintaining the width of all layers under a given maximum width but rely on different strategies to reduce the width of a layer.
The former one yields lower bounds and feasible solutions while the latter can be used to derive upper bounds.

\paragraph{Restricted DDs}

At \cref{restrict-or-relax}, restriction is synonymous with removing surplus nodes.
As detailed by \Cref{alg:layer-reduce}, the least promising nodes of the layer are selected according to a heuristic and simply dropped before resuming the compilation as normal.
As a result, some solutions will not appear in restricted DDs.
Nonetheless, all remaining solutions are feasible.
For a restricted DD $\rst{B}$, we thus have $Sol(\rst{B}) \subseteq Sol(\pb{P})$ and $v(p) = f(x(p)), \forall p \in \rst{B}$.

\begin{algorithm}[t]
\begin{algorithmic}[1]
\WHILE{$\lvert L_j' \rvert > W$}
    \STATE $\mathcal{M} \gets$ select nodes from $L_j'$
    \STATE $L_j' \gets L_j' \setminus \mathcal{M}$
    \vspace{-5pt}
    \begin{framed}
    \vspace{2pt}
    \STATE create node $\mu$ with state $\sigma(\mu) = \oplus(\sigma(\mathcal{M}))$ and add it to $L_j'$ \COMMENT{for relaxation only} \label{relax-start}
    \FORALL{$u \in \mathcal{M}$ \textbf{and} arc $a=\arc[d]{u'}{u}$ incident to $u$}
        \STATE replace $a$ by $a' = \arc[d]{u'}{\mu}$ and set $v(a') = \Gamma_{\mathcal{M}}(v(a),u)$ \label{replace-arc}
    \ENDFOR  \label{relax-end}
    \vspace{2pt}
    \end{framed}
\ENDWHILE
\end{algorithmic}
\caption{Restriction or relaxation of layer $L_j'$ with maximum width $W$.}
\label{alg:layer-reduce}
\end{algorithm}

\paragraph{Relaxed DDs}

As opposed to restricted DDs that delete a part of the solutions, relaxed DDs will never remove feasible solutions but they might introduce infeasible ones.
This is achieved by \textit{merging} nodes together and thus redirecting arcs pointing to different nodes to a single \textit{meta}-node, as performed by \crefrange{relax-start}{replace-arc} of \Cref{alg:layer-reduce}.
It requires defining problem-specific merging operators to merge the corresponding DP states.
If $\mathcal{M}$ is the set of nodes to merge and $\sigma(\mathcal{M}) = \set{\sigma(u) \mid u \in \mathcal{M}}$ the corresponding set of states, the operator $\oplus(\sigma(\mathcal{M}))$ gives the state of the merged node.
The resulting state should encompass all merged states and preserve all their outgoing transitions.
A second operator denoted $\Gamma_\mathcal{M}$ can be specified to adjust the value of the arcs incident to the merged node at \cref{replace-arc}.
As the merging operator gives an approximate representation of all merged states, it can introduce infeasible outgoing transitions.
In addition, merging nodes $u_1,u_2$ allows combining any $\pth{r}{u_1}$ path with any $\pth{u_2}{t}$ path and vice-versa.
Given a valid relaxation operator, we can write for any relaxed DD $\rlx{B}$ that $Sol(\rlx{B}) \supseteq Sol(\pb{P})$ and $v(p) \ge f(x(p)), \forall p \in \rlx{B}$.
To correctly explore the search space, we distinguish \textit{exact nodes} from \textit{relaxed nodes}.
A node $u$ in a DD $\dd{B}$ rooted at node $u_r$ is said \textit{exact} if, for any $\pth{r}{\pth{u_r}{u}}$ path in $\dd{B}$, applying all transitions specified by its arcs recursively from the root leads to the state $\sigma(u)$.
All nodes that do not meet this criterion are called \textit{relaxed}.

\begin{figure}[t]
    \centering
    \input{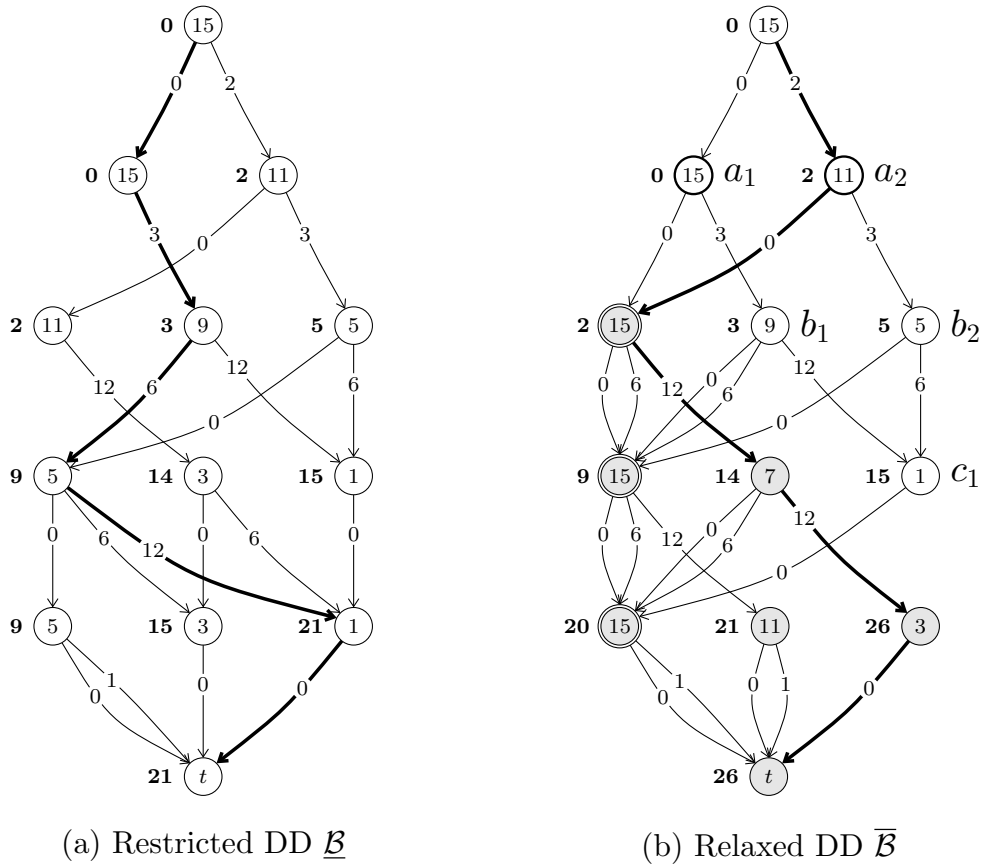}
    \caption{(a) A restricted and (b) a relaxed DD for the BKP instance of \Cref{fig:exact-dd}(a) with $W=3$.
    Merged nodes are circled twice and relaxed nodes are colored in gray.}
    \label{fig:approx-dds}
\end{figure}

\begin{example}
\label{ex:approx-dds}
A valid merging operator for the BKP can simply keep the maximum remaining capacity among the states to merge: $\oplus(\sigma(\mathcal{M})) = \max_{u \in \mathcal{M}} \sigma(u)$.
This ensures that all solutions are preserved since it relaxes the capacity constraint.
The operator $\Gamma_\mathcal{M}$ can be defined as the identity function since there is no need to modify the arc values.
\Cref{fig:approx-dds} shows (a) a restricted and (b) a relaxed DD of width 3 for the BKP instance of \Cref{fig:exact-dd}(a).
In both cases, the nodes $u \in \mathcal{M}$ with the lowest longest $\pth{r}{u}$ path values were selected for removal or merging.
The best solution in the restricted DD gives a lower bound of $21$ while the relaxed DD gives an upper bound of $26$.
This upper bound corresponds to the solution $x = \lst{1,0,2,2,0}$ with total weight $16$, which violates the capacity constraint.
\end{example}

Before explaining how approximate DDs are embedded in a B\&B scheme, let us provide some additional elements of notation.
We denote by $\opt{p}(\pth{u_1}{u_2} \mid \dd{B})$ an optimal path between nodes $u_1$ and $u_2$ in a DD $\dd{B}$ and by $\opt{v}(\pth{u_1}{u_2} \mid \dd{B})$ its value.
For a DD $\dd{B}$ rooted at a node $u_r$, we assume that an exact $\pth{r}{u_r}$ path $p(u_r) := \opt{p}(\pth{r}{u_r} \mid \dd{B}')$ was found and attached to $u_r$ during the prior compilation of another DD $\dd{B}'$.
For conciseness, we denote by $\opt{p}(u \mid \dd{B}) = p(u_r) \cdot \opt{p}(\pth{u_r}{u} \mid \dd{B})$ an optimal path connecting the root node $r$ of the problem and node $u$ within the DD $\dd{B}$, with $\cdot$ the concatenation operator.
The value of this path is given by $\opt{v}(u \mid \dd{B}) = v(p(u_r) \cdot \opt{p}(\pth{u_r}{u} \mid \dd{B})) = v(p(u_r)) + \opt{v}(\pth{u_r}{u} \mid \dd{B})$.
Furthermore, we write $\opt{p}(\dd{B}) = \opt{p}(t \mid \dd{B}), \opt{v}(\dd{B}) = \opt{v}(t \mid \dd{B})$ and $\opt{x}(\dd{B}) = x(\opt{p}(\dd{B}))$ to refer to one of the longest $\pth{r}{t}$ paths in $\dd{B}$, its value and the corresponding variable assignment.
Finally, we define the \textit{successors} of a node $u \in \dd{B}$ as the set of nodes reachable from $u$ in $\dd{B}$, including $u$ itself: $Succ(u \mid \dd{B}) = \set{u} \cup \set{u' \mid (\pth{u}{u'}) \in \dd{B}}$.

\paragraph{A note about reduced DDs} A DD is \textit{reduced} \citep{bryant1986graph,wegener2000} if it does not contain any \textit{equivalent} subgraphs, that is, isomorphic subgraphs where corresponding nodes belong to the same layer and corresponding arcs share the same labels.
DD reduction is very important for many applications, since it produces a unique DD of minimal size for a given variable ordering.
However, \textit{weighted} DDs used for optimization are less amenable to reduction since arc values further restrict which arcs can be superimposed.
\cite{hooker2013decision} explained how transition cost functions of DP models can be rearranged to allow compiling reduced weighted DDs.
Yet, this transformation is not always possible nor necessary since DDs are already reduced in some sense, indeed, they are derived from DP models that superimpose equivalent subproblems by design.

\begin{algorithm}[t]
\begin{algorithmic}[1]
\STATE $Fringe \gets \set{r}$ \COMMENT{a priority queue of nodes ordered by decreasing $v(u) + \ub{v}(u)$}
\vspace{-2pt}
\begin{important}
\STATE $Cache \gets \emptyset$ \COMMENT{a hash table of states to threshold $\sigma(u) \rightarrow \vct{\theta,expanded}$}
\end{important}
\STATE $\lb{x} \gets \bot, \lb{v} \gets -\infty$ \COMMENT{incumbent solution and its value}
\WHILE{$Fringe$ is not empty} \label{b-and-b-loop}
    \STATE $u \gets$ best node from $Fringe$, remove it from $Fringe$ \label{fringe-pop}
    \IF{$v(u) + \ub{v}(u) \le \lb{v}$} \label{prune-bound-2}
        \STATE \textbf{continue}
    \ENDIF
    \begin{important}
    \IF{$Cache.contains(\sigma(u))$} \label{b-and-b-check-cache}
        \STATE $\vct{\theta,expanded} \gets Cache.get(\sigma(u))$
        \IF{$v(u) < \theta$ \textbf{or} $(v(u) = \theta \land expanded)$}
            \STATE \textbf{continue}
        \ENDIF
    \ENDIF
    \end{important}
    \STATE $\rst{B} \gets Restricted(u)$ \label{compute-restricted} \COMMENT{compile restricted DD with \Cref{alg:dd-compilation}}
    \IF[update incumbent]{$\opt{v}(\rst{B}) > \lb{v}$} \label{update-best-start}
        \STATE $\lb{x} \gets \opt{x}(\rst{B}), \lb{v} \gets \opt{v}(\rst{B})$ \label{update-best-end}
    \ENDIF
    \IF{$\rst{B}$ is not exact}
        \STATE $\rlx{B} \gets Relaxed(u)$ \label{compute-relaxed} \COMMENT{compile relaxed DD with \Cref{alg:dd-compilation}}
        \STATE compute LocBs with \Cref{compute-local-bounds} applied to $\rlx{B}$
        \begin{important}
        \STATE update $Cache$ with \Cref{alg:compute-theta} applied to $\rlx{B}$
        \end{important}
        \FORALL{$u' \in EC(\rlx{B})$}
            \STATE $v(u') \gets \opt{v}(u' \mid \rlx{B}), \ub{v}(u') \gets \ub{v}(u' \mid \rlx{B}), p(u') \gets \opt{p}(u' \mid \rlx{B})$ \label{attach-data}
            \IF{$v(u') + \ub{v}(u') > \lb{v}$} \label{prune-bound-1}
                \STATE add $u'$ to $Fringe$ \label{add-fringe}
            \ENDIF
        \ENDFOR
    \ENDIF
\ENDWHILE
\RETURN $(\lb{x}, \lb{v})$
\end{algorithmic}
\caption{The DD-based branch-and-bound algorithm.}
\label{alg:b-and-b}
\end{algorithm}

\subsection{Branch-and-Bound}
\label{section:b-and-b}

\begin{figure}[t]
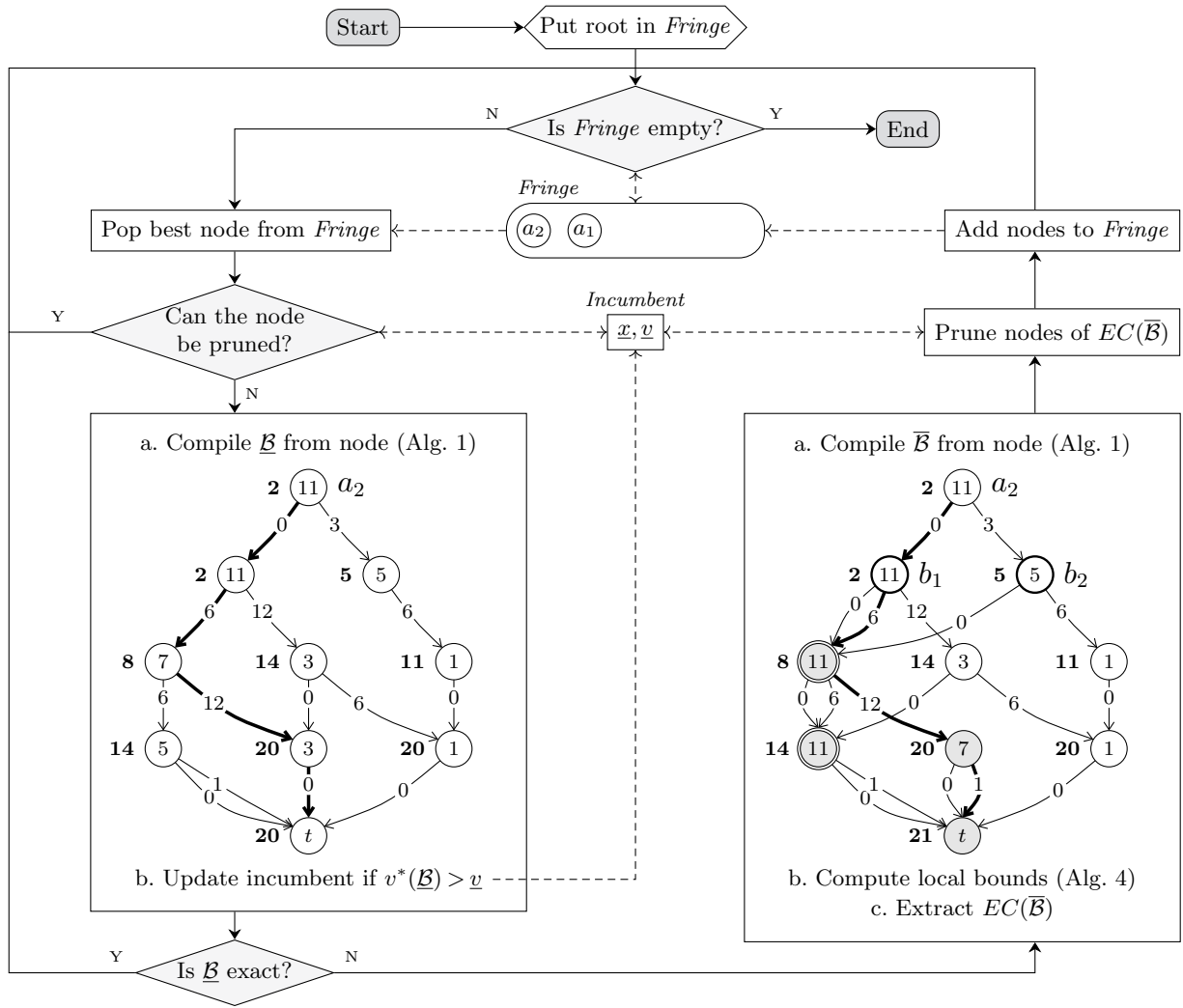

    \centering
    \begin{tikzpicture}[auto,font=\footnotesize,ddnode/.style={circle,draw,inner sep=1pt},relaxed/.style={fill=Black!15},decision/.style={draw,diamond,aspect=3,fill=Black!4},yesno/.style={font=\tiny},flowedge/.style={-{Stealth[length=4pt,width=5pt]}}]

\def\hspaceboxes{7.5}
\def\widthboxes{6}
\def\hspacearrowsa{2}
\def\vspaceboxes{1.4}

\def\hspacelayer{1.5}
\def\vspacelayer{1.2}

\newcommand{\drawnode}[4][]{
    \node[ddnode,#1] (#2) at (#3,0) {#4};
}

\begin{scope}[shift={(0,2*\vspaceboxes)}]
    \node[draw,chamfered rectangle, chamfered rectangle xsep=5pt,inner sep=1pt] (init-fringe) at (0, 0) {Put root in \textit{Fringe}};

    \node[draw,rounded corners=5pt,relaxed] (start) at (-0.5*\hspaceboxes, 0) {Start};
\end{scope}

\begin{scope}[shift={(0,\vspaceboxes)}]
    \node[decision,inner sep=1pt] (check-fringe) at (0, 0) {Is \textit{Fringe} empty?};

    \node[draw,rounded corners=5pt,relaxed] (end) at (0.5*\hspaceboxes, 0) {End};
\end{scope}

\begin{scope} % FRINGE
    \def\fringehspace{0.7}
    \def\fringeradius{0.4}
    
    % NODES
    \drawnode{a2}{-2*\fringehspace}{$a_2$};
    \drawnode{a3}{-1*\fringehspace}{$a_1$};
    \node[ddnode,opacity=0] (fringe-end) at (2*\fringehspace,0) {$a_2$};

    % FRINGE
    \node[draw,rounded corners=10pt,fit={(a2) (fringe-end)}] (fringe) {};
    \node[above right=4pt and 1pt of a2, anchor=south,font=\scriptsize] {\textit{Fringe}};
\end{scope}

\begin{scope}[shift={(-\hspaceboxes, 0)}] % POP NODE
    \node[draw,anchor=west] (pop-node) at (0, 0) {Pop best node from \textit{Fringe}};
\end{scope}

\begin{scope}[shift={(-\hspaceboxes, -\vspaceboxes)}] % CHECK NODE PRUNING
    \node[decision,inner sep=-2pt,anchor=west] (prune-popped-node) at (0, 0) {
        \begin{tabular}{c}
            Can the node\\
            be pruned?
        \end{tabular}};

    \node[decision,inner sep=1pt,below=5.5*\vspaceboxes of prune-popped-node] (continue-exact) {Is $\rst{B}$ exact?};
\end{scope}

\begin{scope}[shift={(0, -\vspaceboxes)}] % INCUMBENT
    \node[draw] (incumbent) at (0, 0) {$\lb{x}, \lb{v}$};
    \node[above=-1pt of incumbent, anchor=south,font=\scriptsize] {\textit{Incumbent}};
\end{scope}

\begin{scope}[shift={(-\hspaceboxes+0.5*\widthboxes, -2.1*\vspaceboxes)}] % RESTRICTED
    \node (compile-restricted) at (0, 0) {a. Compile $\rst{B}$ from node (Alg. 1)};
            
    \tikzset{shift={(0,-4.5*\vspacelayer)}}

    \input{tikz/b-and-b-rst.tikz}
            
    \tikzset{shift={(0,-0.5*\vspacelayer)}}

    \node (update-sol) at (0, 0) {b. Update incumbent if $\opt{v}(\rst{B}) > \lb{v}$};
\end{scope}

\begin{scope}[shift={(\hspaceboxes-0.5*\widthboxes, -2.1*\vspaceboxes)}] % RELAXED
    \node (compile-relaxed) at (0, 0) {a. Compile $\rlx{B}$ from node (Alg. 1)};
            
    \tikzset{shift={(0,-4.5*\vspacelayer)}}

    \input{tikz/b-and-b-rlx.tikz}

    \tikzset{shift={(0,-0.5*\vspacelayer)}}
    
    \node  (compute-locbs) at (0, 0) {b. Compute local bounds (Alg. 4)};
    \node  (extract-cutset) at (0, -0.4) {c. Extract $EC(\rlx{B})$};
\end{scope}

\coordinate (restricted-nw) at (-\hspaceboxes,-1.8*\vspaceboxes) {};
\coordinate (restricted-se) at (-\hspaceboxes+\widthboxes,-6.7*\vspaceboxes) {};
\coordinate (relaxed-nw) at (\hspaceboxes,-1.8*\vspaceboxes) {};
\coordinate (relaxed-se) at (\hspaceboxes-\widthboxes,-7*\vspaceboxes) {};

\node[draw,fit={(restricted-nw) (restricted-se)},inner sep=0] (restricted) {};
\node[draw,fit={(relaxed-nw) (relaxed-se)},inner sep=0] (relaxed) {};

\begin{scope}[shift={(\hspaceboxes, -\vspaceboxes)}] % CHECK CUTSET PRUNING
    \node[draw,anchor=east] (prune-cutset) at (0, 0) {Prune nodes of $EC(\rlx{B})$};
\end{scope}

\begin{scope}[shift={(\hspaceboxes, 0)}] % ENQUEUE CUTSET
    \node[draw,anchor=east] (enqueue-cutset) at (0, 0) {Add nodes to \textit{Fringe}};
\end{scope}

\node (anchor-above-fringe) at (0,1.6*\vspaceboxes) {};
\node (anchor-west) at (-1.15*\hspaceboxes,0) {};
\node (anchor-east) at ( \hspaceboxes-\hspacearrowsa,0) {};

\draw[flowedge] (start) -- (init-fringe);
\draw[-] (init-fringe) -- (anchor-above-fringe.center);
\draw[flowedge] (check-fringe) -- node[yesno,anchor=south,pos=0.05] {N} (check-fringe -| prune-popped-node) -- (pop-node.north -| prune-popped-node);
\draw[flowedge] (check-fringe) -- node[yesno,anchor=south,pos=0.1] {Y} (end);
\draw[<->,densely dashed] (check-fringe.south) -- (fringe.north);
\draw[->,densely dashed] (fringe) -- (pop-node);
\draw[flowedge] (pop-node.south -| prune-popped-node) -- (prune-popped-node.north);
\draw[<->,densely dashed] (prune-popped-node) -- (incumbent);
\draw[flowedge] (prune-popped-node) -- node[yesno,anchor=south,pos=0.4] {Y} (prune-popped-node -| anchor-west) -- (anchor-above-fringe -| anchor-west) -- (anchor-above-fringe -| check-fringe) -- (check-fringe);
\draw[flowedge] (prune-popped-node.south) -- node[yesno,anchor=west,pos=0.4] {N} (restricted.north -| prune-popped-node);
\draw[->,densely dashed] (update-sol) -- (update-sol -| incumbent) -- (incumbent);
\draw[flowedge] (restricted.south -| prune-popped-node) -- (continue-exact.north);
\draw[flowedge] (continue-exact) -- node[yesno,anchor=south,pos=0.025] {N} (continue-exact -| anchor-east) -- (relaxed.south -| anchor-east);
\draw (continue-exact) -- node[yesno,anchor=south,pos=0.15] {Y} (continue-exact -| anchor-west) -- (prune-popped-node -| anchor-west);
\draw[flowedge] (relaxed.north -| anchor-east) -- (prune-cutset.south -| anchor-east);
\draw[<->,densely dashed] (prune-cutset) -- (incumbent);
\draw[flowedge] (prune-cutset.north -| anchor-east) -- (enqueue-cutset.south -| anchor-east);
\draw[->,densely dashed] (enqueue-cutset) -- (fringe);
\draw[-] (enqueue-cutset.north -| anchor-east) -- (anchor-east |- anchor-above-fringe) -- (anchor-above-fringe.center);

\end{tikzpicture}
    \caption{Flowchart of the DD-based B\&B algorithm.}
    \label{fig:b-and-b}
\end{figure}

The two ingredients presented in \Cref{section:dds} allow building a DD-only B\&B algorithm, as introduced by \cite{bergman2016discrete}.
Restricted DDs are used to quickly find quality solutions from any starting node while relaxed DDs recursively decompose the problem at hand and provide bounds for the subproblems thus generated.
The B\&B algorithm is formalized by \Cref{alg:b-and-b} and summarized by \Cref{fig:b-and-b}.
Starting with the root node, \cref{b-and-b-loop} loops over the set of open nodes, contained in a queue referred to as the \textit{Fringe}.
For each of them, a restricted DD $\rst{B}$ is compiled at \cref{compute-restricted}.
It may provide a solution with value $\opt{v}(\rst{B})$ improving the current best $\lb{v}$, in which case the incumbent is updated at \crefrange{update-best-start}{update-best-end}.
If $\rst{B}$ is exact, then this subproblem can be considered as fully explored.
Otherwise, a relaxed DD $\rlx{B}$ must be developed at \cref{compute-relaxed}.
The role of this relaxed DD is twofold.
First, it allows identifying a set of subproblems that need to be further explored so that all parts of the search space are covered.
This set of subproblems is called an \textit{exact cutset} (EC) of $\rlx{B}$.

\begin{definition}[Exact cutset]
\label{def:exact-cutset}
An \textit{exact cutset} $EC(\rlx{B})$ of a relaxed DD $\rlx{B}$ rooted at node $u_r$ is a set of exact nodes such that any $\pth{u_r}{t}$ path crosses at least one of them.
\end{definition}

Thanks to the divide-and-conquer nature of the DP model underlying the whole DD-based B\&B approach, the node $u_r$ corresponds to a subproblem $\pb{P}\lvert_{u_r}$ that can be decomposed in a set of subproblems given by its direct successors in $\rlx{B}$.
By transitivity, it follows that an EC of $\rlx{B}$ contains an exact decomposition of $u_r$, i.e., solving $\pb{P}\lvert_{u_r}$ is equivalent to solving $\pb{P}\lvert_{u}$ for all $u \in EC(\rlx{B})$.
Therefore, nodes of the EC of $\rlx{B}$ are added to the \textit{Fringe} in the B\&B at \cref{add-fringe} in order to pursue the exhaustive exploration of the search space.
The two most commonly used ECs are the \textit{last exact layer} (LEL), which is the deepest layer in $\rlx{B}$ that contains only exact nodes, and the \textit{frontier cutset} (FC) defined as $FC(\rlx{B}) = \set{ u \in \rlx{B} \mid u \text{ is exact } \land \exists a=\arc[d]{u}{u'} \text{ such that } u' \text{ is relaxed}}$.

The second role of relaxed DDs is to compute an upper bound $\ub{v}(u)$ for each node $u \in EC(\rlx{B})$, which is used to discard nodes that are guaranteed to lead to solutions of value worse or equal to $\lb{v}$.
This check is performed at \cref{prune-bound-1} before adding a node to the \textit{Fringe} and is repeated at \cref{prune-bound-2} when a node is selected for exploration.
While the upper bound $\ub{v}(u)$ for each node $u \in EC(\rlx{B})$ was originally based solely on $\opt{v}(\rlx{B})$, stronger pruning can be obtained by using \textit{local bounds} and \textit{rough upper bounds} instead, as detailed in \Cref{additional-bounds}.
In addition to the upper bound $\ub{v}(u)$, \cref{attach-data} also attaches to each cutset node $u$ the longest $\pth{r}{u}$ path $p(u) = \opt{p}(u \mid \rlx{B})$ and its value $v(u) = \opt{v}(u \mid \rlx{B})$.
The cutset nodes are added to the \textit{Fringe} unless their upper bound allows pruning them, and this whole process is repeated until the \textit{Fringe} is emptied.

\begin{example}
Back to our example problem defined by \Cref{fig:exact-dd}(a), the DDs represented in \Cref{fig:approx-dds}(a) and (b) could be respectively the first restricted and relaxed DDs compiled during the B\&B.
The best solution of the restricted DD would thus become the incumbent $\lb{v} = \opt{v}(\rst{B}) = 21$ and an EC should be extracted from the relaxed DD with upper bound $\opt{v}(\rlx{B}) = 26$.
The LEL is the second layer of the relaxed DD, containing only node $a_1$ and $a_2$, while the FC would encompass nodes $a_1, a_2, b_1, b_2$ and $c_1$ since they all have an arc pointing to a relaxed node.
Assuming the LEL is used, \Cref{fig:b-and-b} shows how the B\&B would pursue the search after having added nodes $a_1$ and $a_2$ to the \textit{Fringe} and selecting $a_2$ for exploration.
The restricted DD compiled from $a_2$ fails to improve the incumbent.
Furthermore, the upper bound provided by the relaxed DD is equal to the incumbent value, it is thus unnecessary to enqueue any node of the EC.
\end{example}

\subsection{Additional Filtering Techniques}
\label{additional-bounds}

\cite{gillard2021improving} introduced two additional bounding mechanisms.
The first one is the \textit{rough upper bound} (RUB) that aims at identifying suboptimal nodes as early as during the compilation of approximate DDs.
It does so by computing a cheap problem-specific upper bound $\rub{u}$ for every node at \cref{rub-pruning} of \Cref{alg:dd-compilation}, which can help pruning many nodes before even generating their successors.
The RUB is an additional modeling component that can be specified for the states of a given DP model, such that $\rub{u} \ge \opt{v}(\pth{u}{t} \mid \dd{B})$ for any node $u$ in $\dd{B}$, with $\dd{B}$ the exact DD for problem $\pb{P}$.

In the original DD-based B\&B algorithm, all nodes of the EC retrieved from the relaxed DD developed at \cref{compute-relaxed} of \Cref{alg:b-and-b} are added to the \textit{Fringe} if the single upper bound given by $\opt{v}(\rlx{B})$ is greater than $\lb{v}$.
The \textit{local bounds} (LocBs) presented by \cite{gillard2021improving} refine this reasoning by computing a distinct upper bound for each node $u$ in $\rlx{B}$, given by the value of the longest $\pth{u}{t}$ path in $\rlx{B}$.

\begin{definition}[Local bound]
Given $\rlx{B}$ a relaxed DD for problem $\pb{P}$ and a node $u \in \rlx{B}$, the \textit{local bound} $\locb{u}$ of $u$ within $\rlx{B}$ is given by:
\begin{equation}
\label{equation:local-bound}
    \locb{u} = \left\{ \begin{array}{lcl}
        \opt{v}(\pth{u}{t} \mid \rlx{B}), && \text{if $(\pth{u}{t}) \in \rlx{B}$}, \\
        -\infty, && \text{otherwise.} 
    \end{array} \right.
\end{equation}
\end{definition}

\begin{algorithm}[t]
\begin{algorithmic}[1]
\STATE $i \gets$ index of the root layer of $\rlx{B}$
\STATE $(\fromto{L_i}{L_n}) \gets Layers(\rlx{B})$
\STATE $\locb{u} \gets -\infty, mark(u) \gets$ false \textbf{for each node} $u \in \rlx{B}$ \label{init-locb}
\STATE $\locb{t} \gets 0, mark(t) \gets$ true \label{init-locb-terminal}
\FOR{$j=n$ \textbf{down to} $i$}
    \FORALL{$u \in L_j$}
        \IF{$mark(u)$}
            \FORALL{arc $a=\arc{u'}{u}$ incident to $u$}
                \STATE $\locb{u'} \gets \max \set{\locb{u'}, \locb{u} + v(a)}$  \label{propagate-bottomup-locb}
                \STATE $mark(u') \gets$ true \label{mark-reachable}
            \ENDFOR
        \ENDIF
    \ENDFOR
\ENDFOR
\end{algorithmic}
\caption{Computation of the local bound $\locb{u}$ of every node $u$ in $\rlx{B}$.}
\label{compute-local-bounds}
\end{algorithm}

LocBs can be computed efficiently by performing a bottom-up traversal of $\rlx{B}$, as formalized by \Cref{compute-local-bounds}.
In the algorithm, the $mark$ flags are propagated to all nodes that have at least one path reaching the terminal node $t$.
In parallel, the value $\opt{v}(\pth{u}{t} \mid \rlx{B})$ of each marked node $u$ is computed by accumulating the arc values traversed upwards.
At the terminal node, $\locb{t}$ is set to zero at \cref{init-locb-terminal}.
For all other nodes, the value $\ub{v}_{locb}$ is updated by their direct successors at \cref{propagate-bottomup-locb}.

The upper bound $\ub{v}(u \mid \rlx{B})$ used in \Cref{alg:b-and-b} is obtained by combining the RUB and the LocB derived from the compilation of a relaxed DD $\rlx{B}$:
\begin{equation}
\ub{v}(u \mid \rlx{B}) = \left\{ \begin{array}{lcl}
    \rub{u}, && \text{if $\opt{v}(u \mid \rlx{B}) + \rub{u} \le \lb{v}$,} \\
    \min \set{\rub{u}, \locb{u}}, &&  \text{otherwise.}
\end{array} \right.
\end{equation}
In the case where the RUB caused the node to be pruned at \cref{rub-pruning} of \Cref{alg:dd-compilation}, the value of the RUB is kept.
Otherwise, the minimum of both upper bounds is retained to maximize the pruning potential.
Always keeping the minimum upper bound is also correct, but the definition given simplifies the exposition of the concepts introduced in \Cref{section:cache}.
%The upper bound is then attached to each node $u$ and used to prune nodes at \cref{prune-bound-1,prune-bound-2} of \Cref{alg:b-and-b}.

\begin{figure}
    \centering
    \input{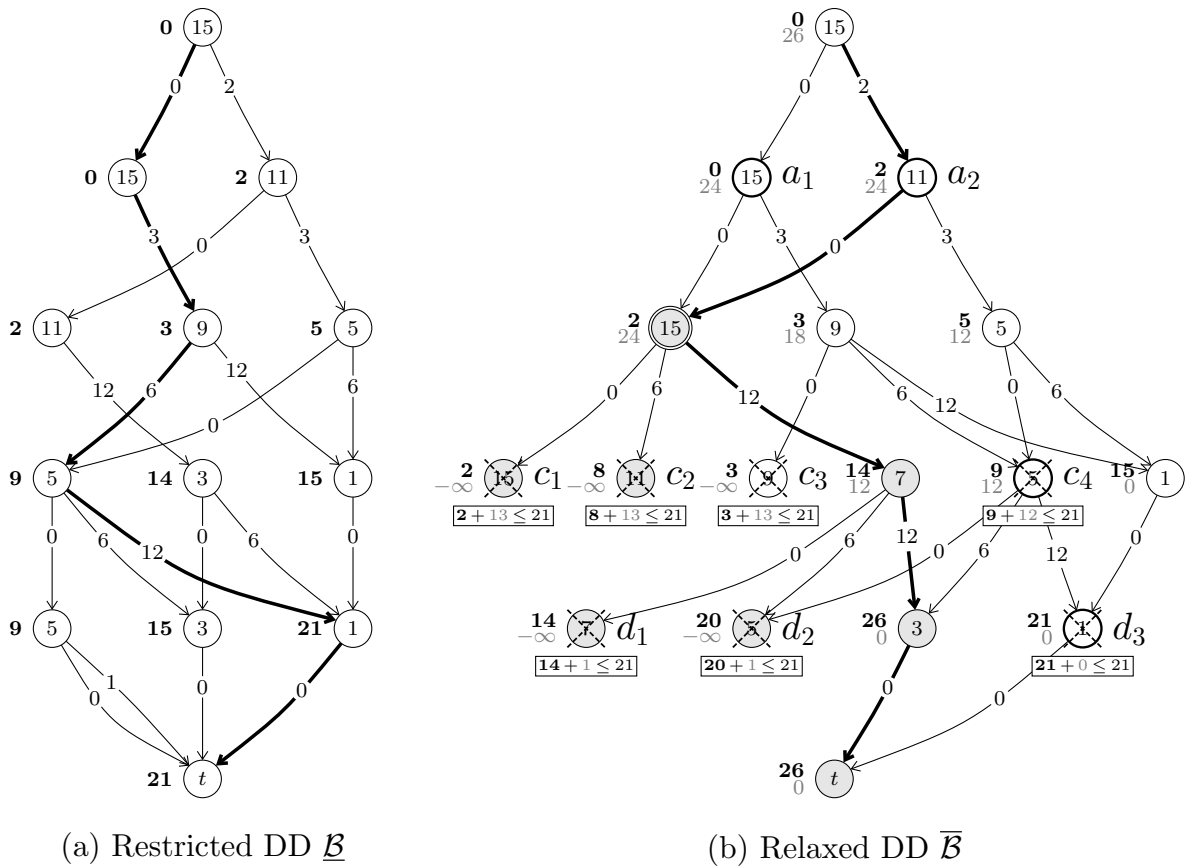}
    \caption{(a) A restricted and (b) a relaxed DD for the BKP instance of \Cref{fig:exact-dd}(a) with $W=3$.
    The LocBs are annotated in gray on the left of each node.
    Pruning decisions are detailed below the filtered nodes.}
    \label{fig:approx-dds-pruning}
\end{figure}

\begin{example}
A naive RUB for the BKP can simply add the maximum quantity of all remaining items to the knapsack, disregarding the capacity constraint.
This can be written as $\ub{v}_{rub}(s^j) = \sum_{k=j}^{n-1} q_k v_k$.
%Of course, a better choice could be to use the LP bound introduced by \cite{dantzig}.
\Cref{fig:approx-dds-pruning} shows approximate DDs for the BKP instance of \Cref{fig:exact-dd}(a) compiled from the root with $W=3$, using RUB and LocB pruning this time.
\Cref{fig:approx-dds-pruning}(a) is the same restricted DD as in \Cref{fig:approx-dds}(a) since RUB pruning requires an initial solution to compare with -- except when infeasibility can be proved.
However, once the lower bound of 21 is obtained, the relaxed DD of \Cref{fig:approx-dds-pruning}(b) can successfully discard nodes $c_1, c_2, c_3, d_1$ and $d_2$ using RUB pruning.
For instance, the RUB for all nodes of the fourth layer is given by $\rub{u} = q_3 v_3 + q_4 v_4 = 2 \times 6 + 1 \times 1 = 13$.
This pruning improves the quality of the relaxed DD as only one layer resorted to node merging, compared to three for the relaxed DD of \Cref{fig:approx-dds}(b).
On the other hand, the LocBs allow filtering nodes $c_4$ and $d_3$ if the FC is used.
Indeed, for each of them we have $\opt{v}(u \mid \rlx{B}) + \locb{u} = 21 \le \lb{v} = 21$.
\end{example}

\section{Caveats of DD-based Branch-and-Bound}
\label{section:caveats}

As explained in \Cref{section:preliminaries}, DD-based B\&B enables solving discrete optimization problems by taking advantage of the compactness of DP models.
Nevertheless, a few observations suggest that all the properties of this type of model have not yet been exploited.
First, we point out that branching does not split the search space into disjoint parts.
Indeed, the very nature of DP models is the ability to solve a large problem by recursively dividing it into smaller \textit{overlapping} subproblems.
Yet, vanilla DD-based B\&B processes subproblems as if they were completely independent.
To give a better intuition of why this might cause the algorithm to waste computational effort, we propose to look at DD-based B\&B as a classical shortest-path algorithm performed on the graph induced by a DP model -- in case of directed acyclic graphs, the shortest-path and longest-path problems are equivalent.
DD-based B\&B can be seen as a combination of \textit{best-first} search at the B\&B level with a sort of \textit{breadth-first} search up to the EC of relaxed DDs, coupled with dedicated bounding and pruning procedures.
However, while it shares many similarities with shortest-path algorithms like A\textsuperscript{*} \citep{hart1968formal}, it lacks one of their important ingredients: the \textit{closed list}.
This data structure collects nodes expanded throughout the search and is used to check whether a given node was already expanded, to avoid expanding it again or adding it to the set of open nodes.

Due to the absence of such data structure, both levels of the B\&B algorithm end up performing some computations multiple times.
At the B\&B level, nothing prevents from adding multiple nodes with the same DP state to the \textit{Fringe} of \Cref{alg:b-and-b} nor triggering the compilation of approximate DDs for several of them at different stages of the search.
Moreover, at the lower level, the approximate DDs can significantly overlap, even when rooted at nodes associated with different DP states.
For instance, suppose we would compute exact DDs for the two subproblems contained in the FC of \Cref{fig:approx-dds-pruning}(b), respectively rooted at $a_1$ and $a_2$.
As one can observe on \Cref{fig:exact-dd}(b), they respectively contain 18 and 15 nodes, among which 11 nodes appear in both DDs.
Therefore, if these two DDs are compiled successively within the B\&B, most of the work done for the first DD is repeated to compile the second one while many transitions could actually yield a value worse or equal than the one obtained before.
In general, the structure of DP state transition systems causes such scenarios to occur very frequently.

The aim of this paper is thus to integrate some form of closed list into DD-based B\&B in order to mitigate the amount of duplicate computations at both levels of the algorithm.
Unfortunately, a simple closed list cannot be used since DP states are not expanded purely in best-first order.
In other words, given a relaxed DD $\rlx{B}$ and an exact node $u \in \rlx{B}$, nothing guarantees that $\opt{p}(u \mid \rlx{B})$ -- the longest path to reach $u$ within $\rlx{B}$ -- is the longest path to reach a node with state $\sigma(u)$ within the complete exact DD.

\section{Branch-and-Bound with Caching}
\label{section:cache}

In this section, we explain how the caveats mentioned in \Cref{section:caveats} can be addressed.
Our idea is to augment DD-based B\&B with a caching mechanism that associates each DP state to an \textit{expansion threshold} that conditions the need for future expansion of nodes with the same DP state.
Formally, after compiling a relaxed DD $\rlx{B}$, an expansion threshold $\theta(u \mid \rlx{B})$ can be derived for each exact node $u \in \rlx{B}$ by exploiting the information contained in $\rlx{B}$.
It is then stored as a \textit{key-value} pair $\vct{\sigma(u),\theta(u \mid \rlx{B})}$ in an associative array referred to as the \textit{Cache}.
When compiling a subsequent approximate DD $\apx{B}$, if a node $u'$ happens to be generated such that $\sigma(u')$ corresponds to an existing key in the \textit{Cache}, the expansion of node $u'$ is skipped if its longest path value $\opt{v}(u' \mid \apx{B})$ is not strictly greater than the expansion threshold stored.
This filtering is carried out at \crefrange{cache-pruning-start}{cache-pruning-end} of \Cref{alg:dd-compilation}.

Concerning the value of these expansion thresholds, a classical DP caching strategy would keep track of the longest path value obtained for each DP state, i.e., by having $\theta(u \mid \rlx{B}) = \opt{v}(u \mid \rlx{B})$ for each exact node $u \in \rlx{B}$.
In this section, we show how stronger expansion thresholds can be obtained by exploiting all the information contained in relaxed DDs.
To give a better intuition on the origin of the expansion thresholds, we split their computation into two distinct thresholds that each cover a specific scenario requiring the expansion node associated with an already visited DP state.
The \textit{dominance} thresholds detect dominance relations between partial solutions found throughout the search, while the \textit{pruning} thresholds extrapolate successful past pruning decisions.
Dominance thresholds can work alone if the RUB and LocB pruning rules are not used, but must be complemented by pruning thresholds whenever these extensions are involved.
In the general case, they are thus combined into a single expansion threshold that covers both scenarios concomitantly.

After giving a formal definition of the aforementioned thresholds, we explain how they can be computed efficiently for all exact nodes visited during the compilation of relaxed DDs before being stored in the \textit{Cache}.
Finally, we describe how the B\&B and the compilation of approximate DDs can be modified to benefit from the expansion thresholds.

\subsection{Dominance Thresholds}
\label{dominance-thresholds}

\begin{figure}[t]
    \centering
    \input{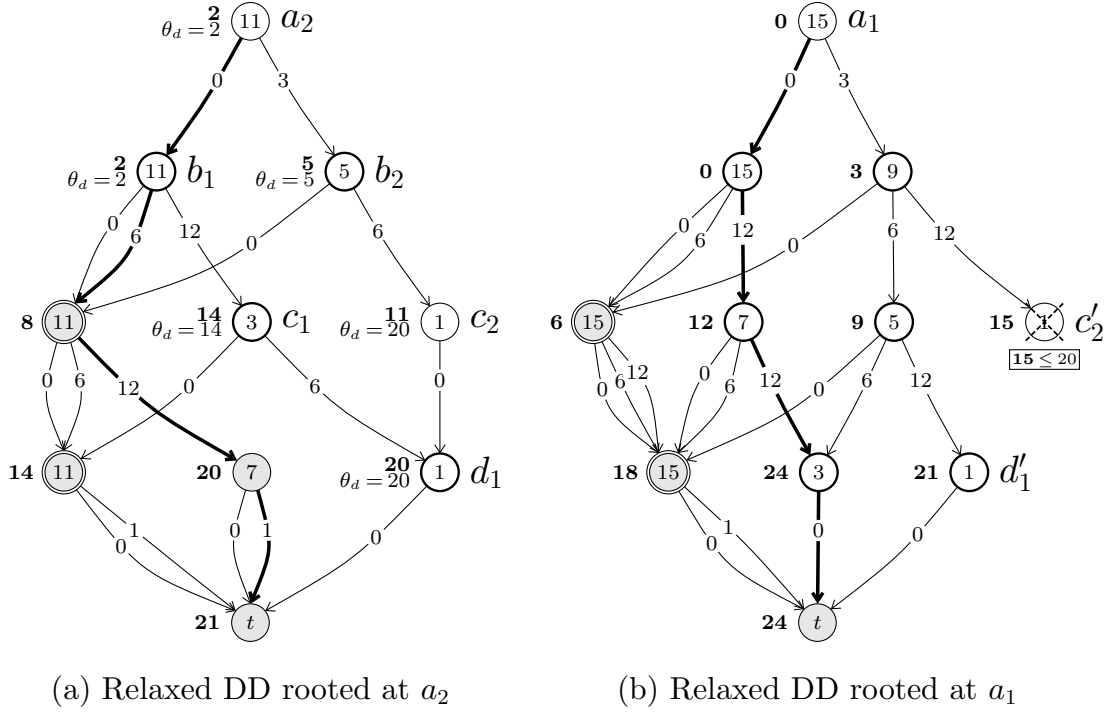}
    \caption{Relaxed DDs with $W=3$ for the BKP instance of \Cref{fig:exact-dd} rooted at nodes (a) $a_2$ and (b) $a_1$.
    Nodes of the relaxed DD (a) are annotated with their dominance threshold, where applicable.
    The relaxed DD (b) is compiled with respect to the dominance thresholds computed in (a).
    }
    \label{fig:dd-with-theta}
\end{figure}

As explained in \Cref{section:b-and-b}, the compilation of a relaxed DD $\rlx{B}$ rooted at node $u_r$ allows decomposing the corresponding subproblem into a set of subproblems given by nodes in $EC(\rlx{B})$.
By extension, any exact node $u_1 \in \rlx{B}$ is also decomposed into a set of subproblems, given by nodes in $EC(\rlx{B})$ that belong to its successors $Succ(u_1 \mid \rlx{B})$.
For example, an exact decomposition of node $c_2$ of \Cref{fig:dd-with-theta}(a) is given by $Succ(c_2 \mid \rlx{B}) \cap EC(\rlx{B}) = \set{d_1}$.
Among these nodes, only those that pass the pruning test at \cref{prune-bound-1} of \Cref{alg:b-and-b} are then added to the \textit{Fringe} of the B\&B.
For simplicity, let us denote this set of nodes by $Leaves_d(u_1 \mid \rlx{B}) = \set{u \in Succ(u_1 \mid \rlx{B}) \cap EC(\rlx{B}) \mid \opt{v}(u \mid \rlx{B}) + \ub{v}(u \mid \rlx{B}) > \lb{v}}$ where $\lb{v}$ is the value of the incumbent solution.
Because we are solving optimization problems, we are only interested in one of the longest $\pth{r}{u_2}$ paths for each node $u_2 \in Leaves_d(u_r \mid \rlx{B})$, and other paths with lower or equal value can safely be ignored.
As a result, we can infer \textit{dominance} relations between the paths obtained within relaxed DDs, and use them later to detect non-dominated paths that are thus still relevant to find the optimal solution of the problem.

\begin{definition}[Dominance]
\label{def:dominance}
Given two $\pth{r}{u}$ paths $p_1$ and $p_2$ in a relaxed DD $\rlx{B}$, $p_1$ is said to dominate $p_2$ -- formally denoted $p_1 \succ p_2$ -- if and only if $v(p_1) > v(p_2)$.
If $v(p_1) \ge v(p_2)$, then $p_1$ weakly dominates $p_2$, expressed as $p_1 \succeq p_2$.
\end{definition}

Given a relaxed DD $\rlx{B}$ and three exact nodes $u_1 \in \rlx{B}$, $u_2 \in Leaves_d(u_1 \mid \rlx{B})$ and $u_1' \notin \rlx{B}$ such that $\sigma(u_1') = \sigma(u_1)$.
The first scenario that requires the expansion of node $u_1'$ happens when it is possible that the concatenation $p_1 \cdot p_2$ of a prefix path $p_1: \pth{r}{u_1'}$ with a suffix path $p_2: \pth{u_1}{u_2}$ dominates the longest path $\opt{p}(u_2 \mid \rlx{B})$ found for the cutset node $u_2$ during the compilation of $\rlx{B}$.
Formally, the dominance relation imposes that $u_1'$ be expanded if $p_1 \cdot p_2 \succ \opt{p}(u_2 \mid \rlx{B})$.
This property leads to the definition below.

\begin{definition}[Individual dominance threshold]
\label{def:individual-dominance-threshold}
Given a relaxed DD $\rlx{B}$ and two exact nodes $u_1 \in \rlx{B}$, $u_2 \in Leaves_d(u_1 \mid \rlx{B})$, the \textit{individual dominance threshold} of $u_1$ with respect to $u_2$ is defined by:
\begin{align}
\theta_{id}(u_1\mid u_2, \rlx{B}) &= \opt{v}(u_2 \mid \rlx{B}) - \opt{v}(\pth{u_1}{u_2} \mid \rlx{B}) \label{eq:individual-dominance-threshold} \\
&= \theta_{id}(u_2 \mid u_2, \rlx{B}) - \opt{v}(\pth{u_1}{u_2} \mid \rlx{B}). \label{eq:individual-dominance-threshold-alt}
\end{align}
\end{definition}

\begin{proposition}
\label{prop:individual-dominance-threshold}
Given a relaxed DD $\rlx{B}$, exact nodes $u_1 \in \rlx{B}$, $u_2 \in Leaves_d(u_1 \mid \rlx{B})$, $u_1' \notin \rlx{B}$ such that $\sigma(u_1') = \sigma(u_1)$, and a path $p_1: \pth{r}{u_1'}$, if 
$v(p_1) > \theta_{id}(u_1\mid u_2, \rlx{B})$ then $p_1 \cdot \opt{p}(\pth{u_1}{u_2} \mid \rlx{B}) \succ \opt{p}(u_2 \mid \rlx{B})$.
Inversely, if $v(p_1) \le \theta_{id}(u_1\mid u_2, \rlx{B})$ then $p_1 \cdot \opt{p}(\pth{u_1}{u_2} \mid \rlx{B}) \preceq \opt{p}(u_2 \mid \rlx{B})$.

\begin{myproof}
By \Cref{eq:individual-dominance-threshold}, we have $v(p_1) > \theta_{id}(u_1\mid u_2, \rlx{B}) = \opt{v}(u_2 \mid \rlx{B}) - \opt{v}(\pth{u_1}{u_2} \mid \rlx{B})$, or equivalently $v(p_1) + \opt{v}(\pth{u_1}{u_2} \mid \rlx{B}) > \opt{v}(u_2 \mid \rlx{B})$.
Using the path value definition, we obtain $v(p_1) + v(\opt{p}(\pth{u_1}{u_2} \mid \rlx{B})) > v(\opt{p}(u_2 \mid \rlx{B}))$.
By concatenating the paths $p_1$ and $\opt{p}(\pth{u_1}{u_2} \mid \rlx{B})$, the inequality becomes $v(p_1 \cdot \opt{p}(\pth{u_1}{u_2} \mid \rlx{B})) > v(\opt{p}(u_2 \mid \rlx{B}))$, which by \Cref{def:dominance} is equivalent to $p_1 \cdot \opt{p}(\pth{u_1}{u_2} \mid \rlx{B}) \succ \opt{p}(u_2 \mid \rlx{B})$.
The proof of the second implication is obtained by replacing each occurrence of $>$ and $\succ$ respectively by $\le$ and $\preceq$ in the proof above.
As $\opt{v}(\pth{u_2}{u_2} \mid \rlx{B}) = 0$, we also trivially have that \Cref{eq:individual-dominance-threshold} and \Cref{eq:individual-dominance-threshold-alt} are equivalent.
\end{myproof}
\end{proposition}

Expanding $u_1'$ is required when a $\pth{r}{u_1'}$ path satisfies the first condition of \Cref{prop:individual-dominance-threshold}, with respect to any node $u_2 \in Leaves_d(u_1 \mid \rlx{B})$.
Therefore, we define a \textit{dominance threshold} for each exact node $u_1 \in \rlx{B}$ that simply computes the least individual dominance threshold with respect to nodes in $Leaves_d(u_1 \mid \rlx{B})$.

\begin{definition}[Dominance threshold]
\label{def:dominance-threshold}
Given a relaxed DD $\rlx{B}$, the \textit{dominance threshold} of an exact node $u_1 \in \rlx{B}$ is given by:
\begin{equation}
\theta_d(u_1 \mid \rlx{B}) = \left\{ \begin{array}{lcl}
    \infty, && \text{if $Leaves_d(u_1 \mid \rlx{B}) = \emptyset$,} \\
    \min_{u_2 \in Leaves_d(u_1 \mid \rlx{B})} \theta_{id}(u_1 \mid u_2, \rlx{B}), && \text{otherwise.}
\end{array}\right.
\end{equation}
\end{definition}

\begin{example}
\label{ex:theta}
Let us illustrate the computation of the dominance thresholds with \Cref{fig:dd-with-theta}(a), showing the relaxed DD compiled from node $a_2$ of our running example.
Note that the dominance thresholds can be used alone here because RUBs and LocBs are disabled, pruning thresholds would otherwise be needed to obtain correct expansion thresholds.
The FC of this relaxed DD consists of the nodes $b_1,b_2,c_1$ and $d_1$.
By applying \Cref{def:dominance-threshold}, we have for each of them that $\theta_d(u \mid \rlx{B}) = \theta_{id}(u \mid u,\rlx{B}) = \opt{v}(u \mid \rlx{B})$.
The dominance thresholds of all other exact nodes in the diagram can be obtained by bottom-up propagation.
For node $c_2$, we obtain $\theta_d(c_2 \mid \rlx{B}) = \theta_{id}(c_2 \mid d_1,\rlx{B}) = \theta_{id}(d_1 \mid d_1,\rlx{B}) - v(c_2 \rightarrow d_1) = 20 - 0 = 20$.

Given those dominance thresholds, \Cref{fig:dd-with-theta}(b) shows the relaxed DD that would be compiled from node $a_1$.
Two pairs of nodes share the same DP state: $c_2,c_2'$ and $d_1,d_1'$.
In (b), node $c_2'$ is pruned since its value is lower than the dominance threshold computed for $c_2$ in (a), even though $c_2'$ has a greater value than $c_2$.
On the other hand, the dominance threshold computed in (a) for $d_1$ is not high enough to prevent the expansion of $d_1'$ in (b).
\end{example}

\subsection{Pruning Thresholds}
\label{pruning-thresholds}

The second contribution concerns nodes that were pruned either because of their RUB or their LocB.
It aims at identifying cases where a new path could circumvent a pruning decision that was made during the compilation of a relaxed DD $\rlx{B}$.
For that purpose, we define a \textit{pruning threshold} for each node $u_1 \in \rlx{B}$.
Similarly to what was done for the dominance thresholds, we denote by $Leaves_p(u_1 \mid \rlx{B})$ the set of successor nodes of $u_1$ that have been pruned because of their RUB or their LocB, i.e., $Leaves_p(u_1 \mid \rlx{B}) = \set{u_2 \in Succ(u_1 \mid \rlx{B}) \mid \opt{v}(u_2 \mid \rlx{B}) + \ub{v}(u_2 \mid \rlx{B}) \le \lb{v}}$ with $\lb{v}$ the value of the incumbent solution at that point.
Formally, given a relaxed DD $\rlx{B}$ and three exact nodes $u_1 \in \rlx{B}$, $u_2 \in Leaves_d(u_1 \mid \rlx{B})$ and $u_1' \notin \rlx{B}$ with $\sigma(u_1') = \sigma(u_1)$, the expansion of node $u_1'$ is required when it is possible that the concatenation $p_1 \cdot p_2$ of a new prefix path $p_1: \pth{r}{u_1'}$ with a suffix path $p_2: \pth{u_1}{u_2}$ obtains a value sufficient to overcome the past pruning decision about node $u_2$.
In other words, the pruning rule imposes that $u_1'$ be expanded if $v(p_1 \cdot p_2) + \ub{v}(u_2 \mid \rlx{B}) > \lb{v}$.
In line with the methodology used for the dominance thresholds, we thus define the \textit{individual pruning threshold} of an exact node $u_1 \in \rlx{B}$ with respect to a node $u_2 \in Leaves_p(u_1 \mid \rlx{B})$.

\begin{definition}[Individual pruning threshold]
\label{def:individual-pruning-threshold}
Given a lower bound $\lb{v}$, a relaxed DD $\rlx{B}$, an exact node $u_1 \in \rlx{B}$ and a pruned node $u_2 \in Leaves_p(u_1 \mid \rlx{B})$, the \textit{individual pruning threshold} of $u_1$ with respect to $u_2$ is defined by:
\begin{align}
\theta_{ip}(u_1 \mid u_2, \rlx{B}) &= \lb{v} - \ub{v}(u_2 \mid \rlx{B}) - \opt{v}(\pth{u_1}{u_2} \mid \rlx{B}) \label{eq:individual-pruning-threshold}\\
&= \theta_{ip}(u_2 \mid u_2, \rlx{B}) - \opt{v}(\pth{u_1}{u_2} \mid \rlx{B}). \label{eq:individual-pruning-threshold-alt}
\end{align}
\end{definition}

\begin{proposition}
\label{prop:individual-pruning-threshold}
Given a lower bound $\lb{v}$, a relaxed DD $\rlx{B}$, an exact node $u_1 \in \rlx{B}$, a pruned node $u_2 \in Leaves_p(u_1 \mid \rlx{B})$, an exact node $u_1' \notin \rlx{B}$ such that $\sigma(u_1') = \sigma(u_1)$, and a path $p_1: \pth{r}{u_1'}$.
If $v(p_1) > \theta_{ip}(u_1\mid u_2, \rlx{B})$ then $v(p_1 \cdot \opt{p}(\pth{u_1}{u_2} \mid \rlx{B})) + \ub{v}(u_2 \mid \rlx{B}) > \lb{v}$.
Inversely, if $v(p_1) \le \theta_{ip}(u_1\mid u_2, \rlx{B})$ then $v(p_1 \cdot \opt{p}(\pth{u_1}{u_2} \mid \rlx{B})) + \ub{v}(u_2 \mid \rlx{B}) \le \lb{v}$.

\begin{myproof}
By \Cref{eq:individual-pruning-threshold}, we have $v(p_1) > \theta_{ip}(u_1\mid u_2, \rlx{B}) = \lb{v} - \ub{v}(u_2 \mid \rlx{B}) - \opt{v}(\pth{u_1}{u_2} \mid \rlx{B})$, or equivalently $v(p_1) + \opt{v}(\pth{u_1}{u_2} \mid \rlx{B}) + \ub{v}(u_2 \mid \rlx{B}) > \lb{v}$.
Using the path value definition, we obtain $v(p_1) + v(\opt{p}(\pth{u_1}{u_2} \mid \rlx{B})) + \ub{v}(u_2 \mid \rlx{B}) > \lb{v}$.
By concatenating the paths $p_1$ and $\opt{p}(\pth{u_1}{u_2} \mid \rlx{B})$, the inequality becomes $v(p_1 \cdot \opt{p}(\pth{u_1}{u_2} \mid \rlx{B})) + \ub{v}(u_2 \mid \rlx{B}) > \lb{v}$.
The proof of the second implication is obtained by replacing each occurrence of $>$ by $\le$ in the proof above.
As $\opt{v}(\pth{u_2}{u_2} \mid \rlx{B}) = 0$, we also trivially have that \Cref{eq:individual-pruning-threshold} and \Cref{eq:individual-pruning-threshold-alt} are equivalent.
\end{myproof}
\end{proposition}

As for the dominance threshold, finding a $\pth{r}{u_1'}$ path satisfying the first condition of \Cref{prop:individual-pruning-threshold} with respect to any node $u_2 \in Leaves_p(u_1 \mid \rlx{B})$ is a sufficient condition to require the expansion of node $u_1'$.
The \textit{pruning threshold} of each exact node $u_1 \in \rlx{B}$ is thus computed as the least individual pruning threshold with respect to any node in $Leaves_p(u_1 \mid \rlx{B})$.

\begin{definition}[Pruning threshold]
\label{def:pruning-threshold}
Given a lower bound $\lb{v}$, a relaxed DD $\rlx{B}$ and an exact node $u_1 \in \rlx{B}$, the \textit{pruning threshold} of $u_1$ within $\rlx{B}$ is given by:
\begin{equation}
\theta_p(u_1 \mid \rlx{B}) = \left\{ \begin{array}{lcl}
    \infty, && \text{if $Leaves_p(u_1 \mid \rlx{B}) = \emptyset$,} \\
    \min_{u_2 \in Leaves_p(u_1 \mid \rlx{B})} \theta_{ip}(u_1 \mid u_2, \rlx{B}), && \text{otherwise.}
\end{array}\right.
\end{equation}
\end{definition}

\begin{figure}[t]
    \centering
    \input{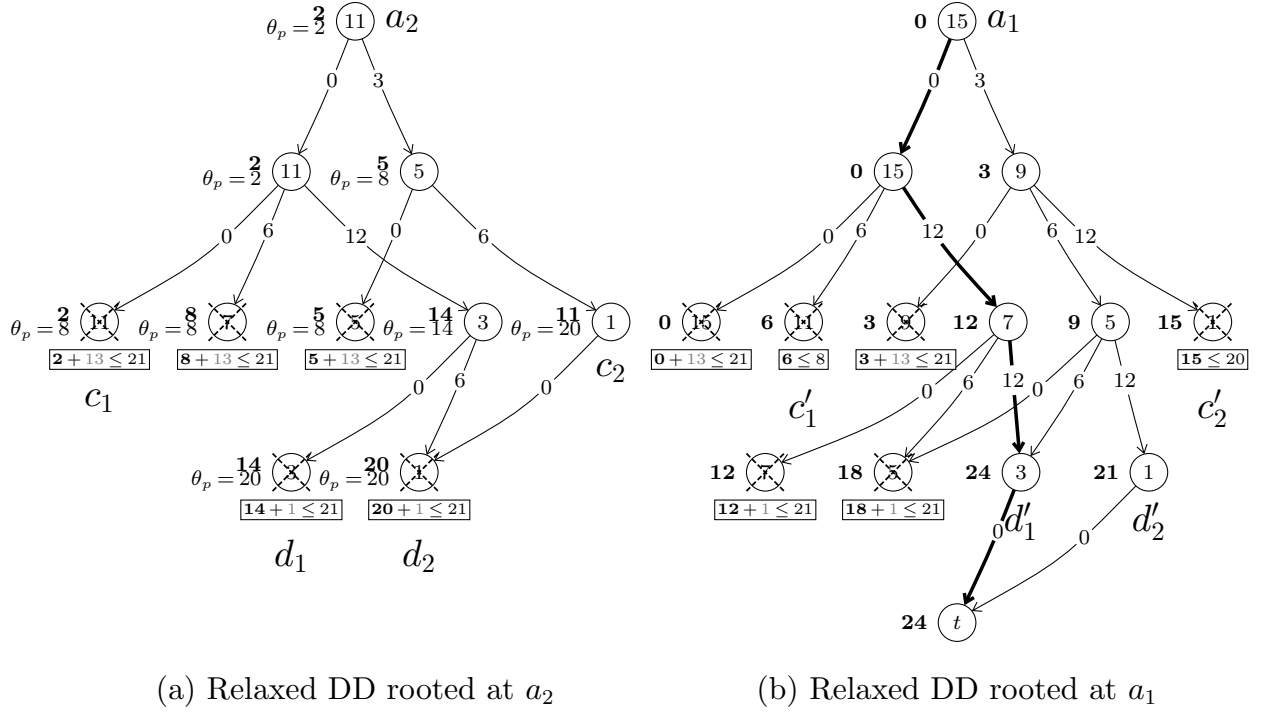}
    \caption{Relaxed DDs with $W=3$ for the BKP instance of \Cref{fig:exact-dd} rooted at nodes (a) $a_2$ and (b) $a_1$.
    Nodes of the relaxed DD (a) are annotated with their pruning threshold.
    The relaxed DD (b) is compiled with respect to the pruning thresholds computed in (a).
    }
    \label{fig:dd-with-theta-pruning}
\end{figure}

\begin{example}
\label{ex:theta-pruning}
We consider the same scenario as in \Cref{ex:theta} of a relaxed DD rooted at node $a_2$, but this time with RUBs and LocBs enabled.
As shown by \Cref{fig:dd-with-theta-pruning}(a), the RUBs manage to prune every path of the DD before reaching the terminal node.
Therefore, we can showcase the pruning thresholds alone, otherwise we should combine them with the dominance thresholds, as will be explained in the next section.
For each pruned node, we have by \Cref{def:pruning-threshold} that $\theta_p(u \mid \rlx{B}) = \theta_{ip}(u \mid u, \rlx{B}) = \lb{v} - \ub{v}(u \mid \rlx{B})$.
For instance, we have $\theta_p(c_1 \mid \rlx{B}) = 21 - 13 = 8$ and $\theta_p(d_2 \mid \rlx{B}) = 21 - 1 = 20$.
We can then propagate these values to compute pruning thresholds of non-pruned nodes such as $c_2$: $\theta_p(c_2 \mid \rlx{B}) = \theta_{ip}(c_2 \mid d_2, \rlx{B}) = \theta_{ip}(d_2 \mid d_2, \rlx{B}) - \opt{v}(\pth{c_1}{d_2} \mid \rlx{B}) = 20 - 0 = 20$.

Once these pruning thresholds are obtained, they can impact the compilation of a relaxed DD from node $a_1$, as illustrated by \Cref{fig:dd-with-theta-pruning}(b).
Among the nodes that share a common DP state across DDs (a) and (b), $c_1'$ and $c_2'$ are successfully filtered by the pruning thresholds computed for $c_1$ and $c_2$, although, like in \Cref{ex:theta-pruning}, the values respectively obtained for $c_1'$ and $c_2'$ in (b) are greater than those of $c_1$ and $c_2$ in (a).
The expansion of nodes $d_1'$ and $d_2'$ cannot be avoided.
However, a new incumbent solution is found after expanding node $d_1'$: $\lb{x} = \lst{0, 0, 2, 2, 0}$ with $\lb{v} = f(\lb{x}) = 24$, which is the optimal solution of the problem.
This concludes the resolution of our BKP instance by the B\&B algorithm since both diagrams of \Cref{fig:dd-with-theta-pruning} are exact and $a_1$ and $a_2$ were the only two nodes in the EC of \Cref{fig:approx-dds-pruning}(b).
\end{example}

\subsection{Expansion Thresholds}

\Cref{dominance-thresholds,pruning-thresholds} presented the two scenarios that necessitate the expansion of nodes associated with a DP state previously reached by an exact node in a relaxed DD, along with the dominance and pruning thresholds used to detect them.
This section explains how these two thresholds are combined to form an \textit{expansion threshold}, and provides a proof of correctness of the accompanying pruning criterion.
It also details the algorithm for computing the expansion threshold for all exact nodes in a relaxed DD.

Dominance and pruning thresholds each give a sufficient condition for the expansion of a node with a previously reached DP state.
The definition of the expansion threshold given below simply ensures that the expansion is triggered if any of those two conditions are met, and inversely that it is avoided if none of those two conditions are met.

\begin{definition}[Expansion threshold]
Given a relaxed DD $\rlx{B}$, the \textit{expansion threshold} of an exact node $u_1 \in \rlx{B}$ is defined by
$\theta(u_1 \mid \rlx{B}) = \min \set{\theta_d(u_1 \mid \rlx{B}), \theta_p(u_1 \mid \rlx{B})}$.
\end{definition}

\begin{proposition}
Given a relaxed DD $\rlx{B}$, an exact node $u_1 \in \rlx{B}$, an exact node $u_1' \notin \rlx{B}$ such that $\sigma(u_1') = \sigma(u_1)$, and a path $p_1 : \pth{r}{u_1'}$,
if $v(p_1) \le \theta(u_1 \mid \rlx{B})$ then $p_1$ can be pruned.

\begin{myproof}
If $v(p_1) \le \theta(u_1 \mid \rlx{B})$ then by definition of the expansion threshold, we have $v(p_1) \le \theta_d(u_1 \mid \rlx{B})$ and $v(p_1) \le \theta_p(u_1 \mid \rlx{B})$.
For the dominance threshold inequality, we have by \Cref{def:dominance-threshold} that $\forall u_2 \in Leaves_d(u_1 \mid \rlx{B}): v(p_1) \le \theta_{id}(u_1 \mid u_2,\rlx{B})$.
When applying \Cref{prop:individual-dominance-threshold}, we obtain $\forall u_2 \in Leaves_d(u_1 \mid \rlx{B}): v(p_1 \cdot \opt{p}(\pth{u_1}{u_2} \mid \rlx{B})) \preceq \opt{p}(u_2 \mid \rlx{B})$.
The second inequality, concerning the pruning threshold, can be developed similarly: by \Cref{def:pruning-threshold}, $\forall u_2 \in Leaves_p(u_1 \mid \rlx{B}): v(p_1) \le \theta_{ip}(u_1 \mid u_2,\rlx{B})$.
And by substituting with \Cref{prop:individual-pruning-threshold}, $\forall u_2 \in Leaves_p(u_1 \mid \rlx{B}): v(p_1\cdot \opt{p}(\pth{u_1}{u_2} \mid \rlx{B})) + \ub{v}(u_2 \mid \rlx{B}) \le \lb{v}$.
Therefore, if $v(p_1) \le \theta(u_1 \mid \rlx{B})$ then the concatenation of $p_1$ with any path $\pth{u_1}{u_2}$ with $u_2 \in Leaves_d(u_1 \mid \rlx{B}) \cup Leaves_p(u_1 \mid \rlx{B})$ is guaranteed to be either weakly dominated or pruned.
Furthermore, by definition of $Leaves_d(u_1 \mid \rlx{B})$ and $Leaves_p(u_1 \mid \rlx{B})$, their union covers all successors of $u_1$ that are in $EC(\rlx{B})$ or that are pruned.
It thus constitutes an exhaustive decomposition of the corresponding subproblem, concluding our argument that $p_1$ is irrelevant for the search.
\end{myproof}
\end{proposition}

In \Cref{def:individual-dominance-threshold,def:individual-pruning-threshold}, one can notice that the individual dominance and pruning thresholds $\theta_{id}(u_1 \mid u_2,\rlx{B})$ and $\theta_{ip}(u_1 \mid u_2,\rlx{B})$ only depend on the value $\opt{v}(\pth{u_1}{u_2} \mid \rlx{B})$, apart from the thresholds $\theta_{id}(u_2 \mid u_2,\rlx{B})$ and $\theta_{ip}(u_2 \mid u_2,\rlx{B})$.
For a relaxed DD rooted at node $u_r$, if those initial individual thresholds are correctly set for nodes in $Leaves_d(u_r \mid \rlx{B})$ and $Leaves_p(u_r \mid \rlx{B})$, the expansion thresholds can thus be computed by performing a single bottom-up pass on the relaxed DD $\rlx{B}$, as described by \Cref{alg:compute-theta}.

\begin{algorithm}[t]
\begin{algorithmic}[1]
\STATE $i \gets$ index of the root layer of $\rlx{B}$
\STATE $(\fromto{L_i}{L_n}) \gets Layers(\rlx{B})$
\STATE $\theta(u \mid \rlx{B}) \gets \infty$ \textbf{for all} $u \in \rlx{B}$ \label{init-theta}
\FOR{$j=n$ \textbf{down to} $i$} \label{bottom-up-loop}
    \FORALL{$u \in L_j$} \label{all-nodes-loop}
        \IF{$Cache.contains(\sigma(u))$ \textbf{and} $\opt{v}(u \mid \rlx{B}) \le \theta(Cache.get(\sigma(u)))$} \label{pruned-by-cache}
            \STATE $\theta(u \mid \rlx{B}) \gets \theta(Cache.get(\sigma(u)))$
        \ELSE
            \IF[rough upper bound pruning]{$\opt{v}(u \mid \rlx{B}) + \rub{u} \le \lb{v}$} \label{pruned-by-bound}
                \STATE $\theta(u \mid \rlx{B}) \gets \lb{v} - \rub{u}$ \label{init-theta-pruned}
            \ELSIF{$u \in EC(\rlx{B})$}
                \IF[local bound pruning]{$\opt{v}(u \mid \rlx{B}) + \locb{u} \le \lb{v}$}
                    \STATE $\theta(u \mid \rlx{B}) \gets \min \set{ \theta(u \mid \rlx{B}), \lb{v} - \locb{u}}$ \label{init-theta-locb}
                \ELSE
                    \STATE $\theta(u \mid \rlx{B}) \gets \opt{v}(u \mid \rlx{B})$ \label{init-theta-fc}
                \ENDIF
            \ENDIF
            \IF{$u$ is exact} \label{cache-insert-check}
                \STATE $expanded \gets$ false \textbf{if} $u \in EC(\rlx{B})$ \textbf{else} true \label{set-expanded-flag}
                \STATE $Cache.insertOrReplace(\sigma(u), \vct{\theta(u \mid \rlx{B}), expanded})$ \label{cache-insert}
            \ENDIF
        \ENDIF
        \FORALL{arc $a=\arc[d]{u'}{u}$ incident to $u$} \label{propagate-theta-start}
            \STATE $\theta(u' \mid \rlx{B}) \gets \min \set{\theta(u' \mid \rlx{B}), \theta(u \mid \rlx{B}) - v(a)}$  \label{propagate-theta-end}
        \ENDFOR
    \ENDFOR
\ENDFOR 
\end{algorithmic}
\caption{Computation of the threshold $\theta(u \mid \rlx{B})$ of every exact node $u$ in a relaxed DD $\rlx{B}$ and update of the \textit{Cache}.}
\label{alg:compute-theta}
\end{algorithm}

The thresholds are first initialized at \cref{init-theta} with a default value of $\infty$ for all nodes.
Then, the loops at \cref{bottom-up-loop,all-nodes-loop} iterate through all nodes of the DD, starting with the nodes of the last layer up to those of the first layer, and propagate threshold values in a bottom-up fashion.
Depending on whether they are pruned or belong to the EC, several nodes require a careful initialization of their expansion threshold.
\begin{itemize}
    \item \textbf{Pruned by the \textit{Cache}:} \cref{pruned-by-cache} checks whether the \textit{Cache} already contains an expansion threshold greater than $\opt{v}(u \mid \rlx{B})$.
    If so, the node was pruned and the previously computed threshold can simply be recycled, as it remains valid throughout the entire algorithm.
    \item \textbf{Pruned by RUB:} when reaching a node $u$ pruned with its RUB during the top-down compilation at \cref{pruned-by-bound}, \cref{init-theta-pruned} sets its expansion threshold $\theta_{ip}(u \mid u, \rlx{B})$ the individual pruning threshold of $u$ relative to itself, because it has no other successor.
    \item \textbf{Pruned by LocB:} if a node $u$ of the EC was pruned because of its LocB however, the expansion threshold is computed at \cref{init-theta-locb} as the minimum between $\theta_{ip}(u \mid u, \rlx{B})$ the individual pruning threshold relative to $u$ and the current value of $\theta(u \mid \rlx{B})$, which accounts for the individual dominance and pruning thresholds relative to successors of $u$.
    \item \textbf{EC node:} the expansion threshold of each cutset node $u$ that will be added to the \textit{Fringe} is simply given by $\theta_{id}(u \mid u, \rlx{B})$ the individual dominance threshold of $u$ relative to itself, as shown at \cref{init-theta-fc}.
\end{itemize}

\Cref{propagate-theta-start,propagate-theta-end} then take care of the bottom-up propagation of the expansion thresholds as to obtain the correct threshold values for all nodes.
The last step at \cref{cache-insert} is to save the expansion threshold computed for each exact node in the \textit{Cache} for later use.
The \textit{Cache} is an associative array storing \textit{key-value} pairs $\vct{\sigma(u), \theta(u \mid \rlx{B})}$, implemented as a \textit{hash table} in practice.
Note that \Cref{alg:compute-theta} can be applied to any type of EC, but an important detail is that thresholds should not be inserted in the \textit{Cache} for exact successors of nodes in the EC to allow for their later expansion.
As the FC contains all the deepest exact nodes, the algorithm can be applied as is.
For the LEL however, a check must be added at \cref{cache-insert-check}.

In addition to the threshold value, a flag called \textit{expanded} is stored in the \textit{Cache} to distinguish two types of expansion thresholds.
First, the nodes that have been added to the \textit{Fringe} from the EC of a relaxed DD $\rlx{B}$ with $\theta(u \mid \rlx{B}) = \opt{v}(u \mid \rlx{B})$ and that need to be further explored by the B\&B, therefore being associated with an expanded flag set to \textit{false} at \cref{cache-insert} of \Cref{alg:compute-theta}.
And second, all exact nodes above the EC of any relaxed DD $\rlx{B}$ in which the threshold $\theta(u \mid \rlx{B})$ was obtained.
In this case, they can be considered as expanded since all their outgoing paths were either pruned or cross a node in the EC of $\rlx{B}$.
When those thresholds are inserted in the \textit{Cache} at \cref{cache-insert} of \Cref{alg:compute-theta}, their \textit{expanded} flag is thus set to $true$.

\subsection{Filtering the Search Using the Cache}

Now that we have presented how expansion thresholds can be computed for all exact nodes of relaxed DDs, we can now elaborate on how these values are utilized within the B\&B.
There are two places where the \textit{Cache} might be beneficial and prune some nodes.
As mentioned before, and shown in \Cref{ex:theta,ex:theta-pruning}, the first is in the top-down compilation of approximate DDs, as described by \Cref{alg:dd-compilation}.
At \cref{forbidden-transition}, the \textit{Cache} is queried and whenever the value $\opt{v}(u \mid \dd{B})$ of a node $u$ fails to surpass the threshold $\theta$, it is added to the $pruned$ set.
This set of pruned nodes is used to define $L_j'$ at \cref{remove-pruned}, a clone of the $j$-th layer from which the pruned nodes have been removed.
In the rest of the algorithm, the pruned layer $L_j'$ is employed instead of $L_j$ to prevent generating any outgoing transition from the pruned nodes.
Nodes from the $pruned$ set are kept in the original layer $L_j$ so that the threshold computations of \Cref{alg:compute-theta} can be performed correctly.

Furthermore, the \textit{Cache} can be used when selecting a node for exploration in the B\&B loop.
At \cref{b-and-b-check-cache} of \Cref{alg:b-and-b}, the \textit{Cache} is queried to potentially retrieve an entry with a threshold value as well as an \textit{expanded} flag.
If a threshold is indeed present, two cases arise depending on the value of the \textit{expanded} flag.
If it is \textit{True}, the node is ignored if its value is less or equal to the threshold.
When \textit{False} however, the node is only ignored if its value is strictly less than the threshold, because in case of equality, this node will be the first to be expanded with that value.
As this filtering happens at the B\&B level, it is not repeated for the root node of approximate DDs, see \cref{ignore-first-layer} of \Cref{alg:dd-compilation}.

An interesting implication of the use of expansion thresholds is that during the bottom-up traversal of relaxed DDs done to compute LocBs, it is possible to detect \textit{dangling nodes}, i.e., nodes with no feasible outgoing transition.
As they have no path to the terminal node, their LocB is $-\infty$ and the expansion threshold stored is $\infty$.
As a result, the bottom-up propagation of the thresholds permits to detect both suboptimality and infeasibility earlier in the future compilations of approximate DDs, in addition to dominance relations discussed before.
%Additionally, filtering relaxed DDs with expansion thresholds strengthens the LocBs further since they are computed only on paths that .
%The LocBs can thus be seen as an estimation .
In terms of complexity, computing the thresholds requires a single traversal of the relaxed DDs compiled, so it has the same complexity as \Cref{alg:dd-compilation} and \Cref{compute-local-bounds} that respectively handle the top-down compilation and the computation of the LocBs.

\section{Limitations}
\label{section:limitations}

The thresholds presented in \Cref{section:cache} offer new pruning perspectives that will be shown to be very impactful in \Cref{section:experiments}.
Yet, they exhibit two main limitations that are discussed in this section.

\subsection{Memory Consumption}

In order to apply B\&B with caching, extra information must be stored in memory.
Some effort is done to reduce memory consumption by deleting thresholds as soon as they are no longer required by the algorithm.
A sufficient condition to remove a threshold is when it concerns a node located in a layer above the \textit{first active layer}.

\begin{definition}[First active layer]
Given $\dd{B}$ the exact DD for problem $\pb{P}$ with layers $\fromto{L_0}{L_n}$.
The \textit{first active layer} of the B\&B specified by \Cref{alg:b-and-b} is defined as the least index $j$ such that a node of layer $L_j$ is in the \textit{Fringe} or is currently selected for exploration at \cref{fringe-pop}.
\end{definition}

Thresholds related to nodes above the first active layer can be safely removed from the \textit{Cache} as there is no way of reaching the associated DP states again.
If memory consumption was nevertheless an issue, a simple solution would be to delete an arbitrary subset of the thresholds stored in the \textit{Cache}.
This does not compromise the optimality guarantees of the algorithm since the only effect of thresholds is to prevent the expansion of nodes associated with already visited DP states.
However, removing some thresholds decreases the pruning perspectives of the algorithm.
Some memory will therefore be saved at the cost of speed.
One could even imagine using eviction rules based on an activity measure of the thresholds in the \textit{Cache}.
Note that all the algorithms presented in the paper are written as if the \textit{Cache} could remove some thresholds along the way.

An argument that could be held against the use of the \textit{Cache} in the context of DD-based B\&B is that in the worst case, it still might need to accommodate as many nodes as there are nodes in the exact DD encoding the problem being solved.
While this argument is true, we would like to point out that the aforementioned explosive worst-case memory requirement already plagued the original B\&B algorithm as it does not implement any measures to limit the size of the \textit{Fringe}.
Furthermore, we would like to emphasize that maintaining the \textit{Cache} is no more costly than maintaining the \textit{Fringe} and significantly less expensive than memorizing an actual instantiation of the exact DD.
Indeed, the maintenance of both the \textit{Cache} and the \textit{Fringe} is $O(|U|)$ -- where $U$ is the set of nodes in the state space -- whereas the requirement to store an actual instantiation of the exact DD in memory is $O(|U| + |A|)$ where $A$ is the set of arcs connecting the nodes in $U$.
It is also worth mentioning that in that case, $|A|$ dominates $|U|$ since by definition of the domains and transition relations, the size of $A$ is bounded by the product of the domain sizes $\Pi_{j=\fromto{0}{n-1}}|D_j|$.

\subsection{Variables Orderings}

Many discrete optimization problems have an imposed variable ordering in their natural DP model.
For instance, DP models for sequencing problems usually make decisions about the $j$-th position of the sequence at the $j$-th stage of the DP model \citep{cire2013multivalued}.
For DP models that allow it, however, it has been shown that variable orderings can yield exact DDs of reduced size as well as approximate DDs with tighter bounds \citep{bergman2012variable,cappart2022improving,karahalios2022variable}.
When DDs are used within a B\&B algorithm, variable orderings thus constitute an additional heuristic that can speed up the search.
They can be separated in two categories: \textit{static} and \textit{dynamic} variable orderings.
The former refers to a single variable ordering used for all DD compilations during the B\&B.
The latter denotes a heuristic that dynamically decides which variable to branch on when generating the next layer of a DD, based on the states contained in the current layer.
Since the techniques introduced in this paper are solely based on the overlapping structure of the DP models, a dynamic variable ordering would most likely compromise much of the expected pruning.
Indeed, it seems unlikely that many states would overlap in DDs compiled with different variable orderings, although it ultimately depends on the modeling of each specific problem.

\section{Computational Experiments}
\label{section:experiments}

This section presents the results of the computational experiments that were conducted to evaluate the impact of the additional pruning techniques presented in this paper.
We performed experiments on three different well-known discrete optimization problems: the asymmetric variant of the Traveling Salesman Problem with Time Windows (TSPTW), the Pigment Sequencing Problem (PSP) -- also known as the Discrete Lot Sizing and Scheduling Problem (DLSP) in the literature -- and the Single-Row Facility Layout Problem (SRFLP).
We describe below the experimental setting used for each problem.

\paragraph{TSPTW}
The solvers were tested on a classical set of benchmark instances introduced in the following papers \citep{ascheuer1996hamiltonian,dumas1995optimal,gendreau1998generalized,langevin1993two,ohlmann2007compressed,pesant1998exact,potvin1996vehicle}.
A dynamic width was used, where the maximum width for layers at depth $j$ is given by $n \times (j+1) \times \alpha$ with $n$ the number of variables in the instance and $\alpha$ a multiplying parameter for which several values were used in the experiments.

\paragraph{PSP}

A set of randomly generated instances was created with the number of items in  $\set{5, 7, 10}$, the number of periods in $\set{50, 100, 150, 200}$ and the density in $\set{0.9, 0.95, 1}$.
The density is computed as the number of demands over the number of time periods.
An additional parameter $\rho \in \set{0.001, 0.01, 0.1}$ was introduced to control the proportion between the stocking costs and the changeover costs.
The costs are then sampled uniformly in $[\rho \cdot 5000, \rho \cdot 15000]$ and $[(1 - \rho) \cdot 5000, (1 - \rho) \cdot 15000]$, respectively.
To generate the demands, item type and time period pairs were selected uniformly among all possible values and added to the instance as long as it remained feasible.
Five instances were generated for each combination of the parameters mentioned.
A fixed width was used but which is also proportional to the number of variables: $n \times \alpha$.

\paragraph{SRFLP} 

The experiments consist of a compilation of benchmark instances from \citep{amaral2006exact,amaral2008exact,amaral2009new,anjos2005semidefinite,anjos2008computing,anjos2009provably,duff1989sparse,heragu1991efficient,hungerlander2013computational,nugent1968experimental,obata1980quadratic,sarker1989amoebic,simmons1969one,yu2003directional} for the SRFLP as well as the Single-Row Equidistant Facility Layout Problem and the Minimum Linear Arrangement Problem -- two of its subproblems.
We limited our experiments to instances with fewer than 50 departments.
The same width strategy as for the PSP was applied.

The models used to encode these problems within the DD framework are described in the Appendix.
For all the experiments, the best results were obtained with an LEL cutset for the classical B\&B (B\&B) and with an FC for B\&B with caching (B\&B+C).
It is worth mentioning that the baseline approach already uses duplicate state detection in the \textit{Fringe}.
Concerning the heuristics of the approach, the node with the highest upper bound is selected first in the B\&B loop.
Moreover, nodes with the smallest longest path value are respectively deleted or merged in restricted and relaxed DDs compilations.
For each problem, we tested multiple maximum widths by varying the parameter $\alpha \in \set{1, 10, 100}$.
The DD approach was implemented in Rust, based on DDO \citep{gillard2021ddo} which is a generic library for DD-based optimization.
The source code and all benchmark instances are available online \citep{DdoCaching}.
In order to showcase the effectiveness of the DD approach, it is compared against a state-of-the-art MIP model for each problem respectively introduced by \citep{hungerlander2018efficient} as Formulation (1) for the TSPTW, by \citep{pochet2006production} as PIG-A-3 for the PSP and by \citep{amaral2009new} for the SRFLP.
The MIP models were solved with Gurobi 9.5.2 \citep{gurobi} with the default settings.
For all problems, the solvers were given 1800 seconds to solve each instance to optimality, using a single thread.

\subsection{Impact of the Caching Mechanism}
\label{results-speed}

\begin{figure}
    \centering
    \includegraphics[width=\textwidth]{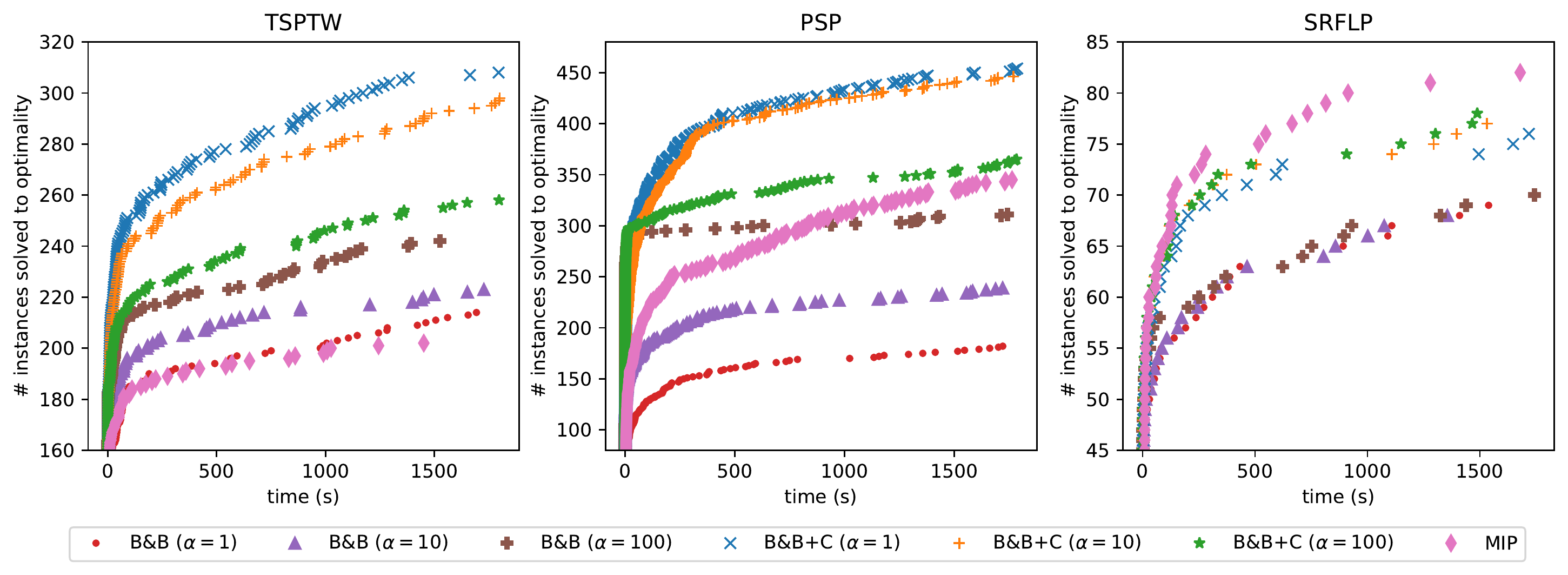}
    \caption{Number of instances solved over time by B\&B, B\&B+C and Gurobi for three different problems. The DD approaches use different maximum widths with $\alpha \in \set{1,10,100}$.}
    \label{fig:results}
    \includegraphics[width=\textwidth]{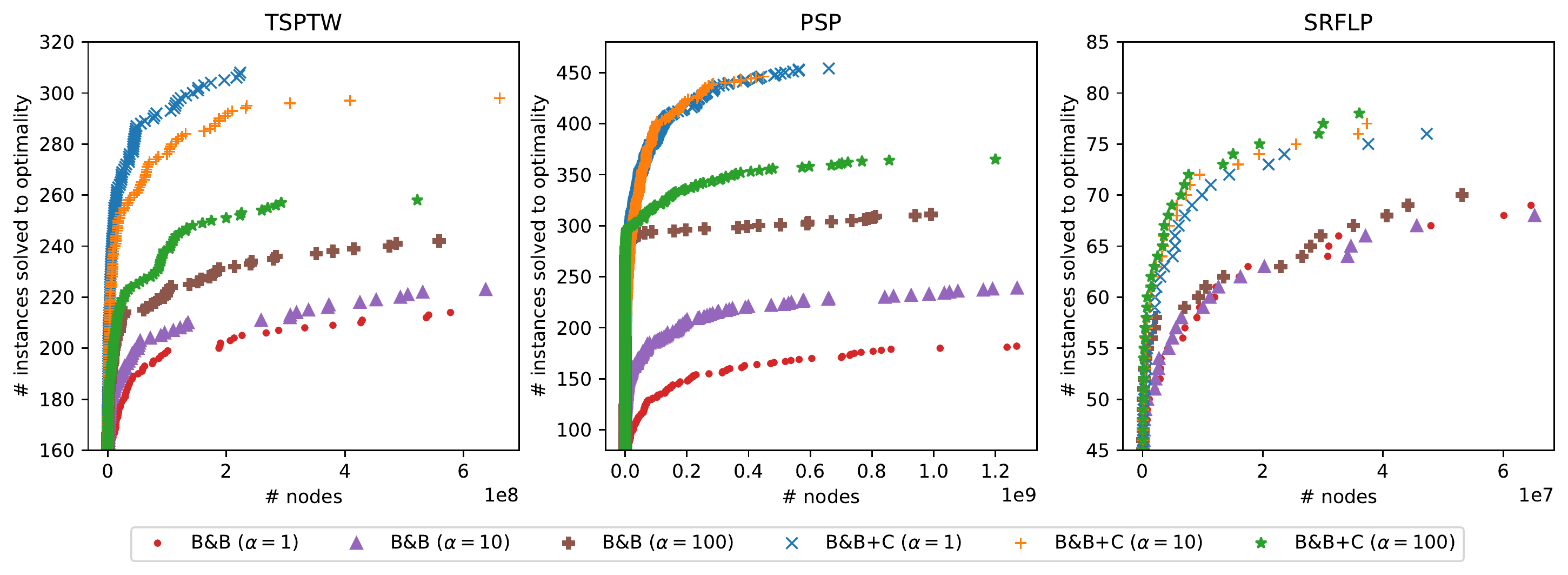}
    \caption{Number of instances solved by B\&B and B\&B+C with respect to the number of DD nodes expanded for three different problems. Different maximum widths were used with $\alpha \in \set{1,10,100}$.}
    \label{fig:results-nodes}
\end{figure}

\Cref{fig:results,fig:results-nodes} show the results of these experiments, respectively in terms of computation time and number of DD nodes expanded during the search.
This measure of nodes expanded accounts for all nodes expanded during top-down compilations of restricted and relaxed DDs.
Each graph presents the total number of instances solved under any given time or nodes expanded.
For each of the studied problems and whatever the maximum width used, B\&B+C is able to solve significantly more instances than B\&B within the time limit.
This speedup is directly linked to a reduction of the number of DD nodes expanded, as can be seen \Cref{fig:results-nodes}.
This confirms our intuition that a lot of work is unnecessarily repeated by B\&B and shows that the pruning techniques introduced in this paper help neutralize much of this problem.
Moreover, if we look at the performance achieved using different maximum widths for the TSPTW and the PSP, one can notice that for B\&B, increasing the width helps to solve more instances.
Indeed, larger DDs allow stronger bounds to be derived and instances to be closed more quickly, as long as they are not too expensive to compile.
Yet, the opposite observation can be made about B\&B+C: using smaller maximum widths results in solving more instances for the TSPTW and the PSP.
This perfectly captures the double benefit of the \textit{Cache}.
The strength of dominance and pruning thresholds allow discarding many transitions during the compilation of approximate DDs and avoid repeating previous work.
As a result, narrower DDs can explore and prune the search space as fast while being much cheaper to generate, and sometimes identify uninteresting parts earlier in the search.
Again, this observation about \Cref{fig:results} can be validated on \Cref{fig:results-nodes} where B\&B+C manages to solve significantly more instances within a given maximum number of nodes when using narrower DDs.
Another benefit of B\&B+C is that it fully exploits the potential of FCs: as all non-improving transitions are blocked by the \textit{Cache}, it is only natural to use the deepest possible exact cutset.
The same cannot be said in the case of B\&B because frontier cutsets usually contain many nodes, some of which are the parents of others, and this only exacerbates the caveats mentioned in \Cref{section:caveats}.

For the SRFLP, the difference of performance obtained by varying the maximum width is not as significant as for the other two problems.
This is probably because the problem is unconstrained and the relaxation is not very tight.
As a result, the thresholds computed with small width DDs may not be strong enough to avoid exploring nodes that are explored by DDs of larger widths.
As a matter of fact, slightly fewer instances are solved using narrow DDs for a same number of DD nodes expanded.
In this case, wider DDs may derive better bounds and thresholds and perform slightly better.
Still, using a \textit{Cache} with DDs of any width helps to close more instances than without it.

To put these results into perspective, we compare them on \Cref{fig:results} to those obtained with MIP models solved with Gurobi.
For the TSPTW, Gurobi could only solve 202 instances, which is less than the worst-performing configuration of DDO.
For the PSP, B\&B+C with $\alpha = 1$ could solve 454 of the generated instances while Gurobi solved 345 of them.
On the other hand, B\&B with $\alpha = 100$ solved only 311 instances.
The addition of the \textit{Cache} thus helped improve DD-based B\&B by a large margin and allowed outperforming Gurobi on this set of instances.
For the SRLFP, B\&B and B\&B+C with $\alpha = 100$ were able to solve respectively 12 and 4 instances less than Gurobi within the given time budget.
Therefore, even if the best results are achieved with the MIP model, using the \textit{Cache} closed much of the gap that separates the two techniques.

\begin{figure}
    \centering
    \includegraphics[width=\textwidth]{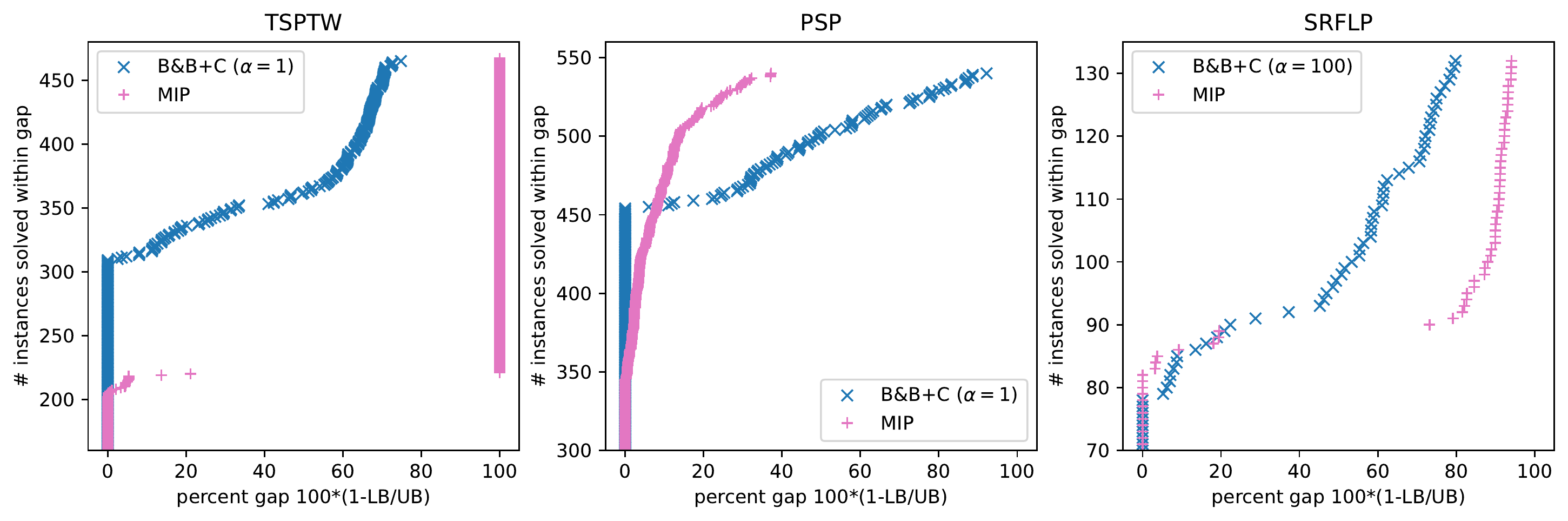}
    \caption{Number of instances solved by the best performing B\&B+C and Gurobi within a given optimality gap for each of the three problems.}
    \label{fig:results-gap}
\end{figure}

Slightly different observations can be made through \Cref{fig:results-gap}, which compares the number of instances solved within a given optimality gap by B\&B+C with the best choice of $\alpha$ for each problem and by Gurobi.
For the TSPTW, it appears that Gurobi fails to find an admissible solution for nearly all unsolved instances, which results in an optimality gap of 100\%.
B\&B+C does not suffer from this problem, yet the optimality gap for unsolved instances is quite large in most cases.
Even though Gurobi solved fewer instances for the PSP, it manages to solve all instances within a 40\% optimality gap and most of them within a 20\% optimality gap.
The optimality gap obtained by B\&B+C for unsolved instances is much poorer.
Surprisingly, B\&B+C achieves tighter optimality gaps for hard SRFLP instances: all of them are within 80\% while for Gurobi, many unsolved instances lie outside this optimality gap.
In general, the optimality gaps obtained for unsolved instances with B\&B+C seem to deteriorate very rapidly as the instances get harder.
One would therefore be tempted to infer that a lot of computational effort is still required to solve the remaining instances -- even those with a relatively small gap.
However, we believe that, in the case of B\&B+C, the optimality gap and the remaining computational effort are not perfectly correlated.
Indeed, having filled the \textit{Cache} with thresholds makes it easier to process the remaining nodes.
%, even though their upper bound was not explicitly tightened.
%This is precisely what the peeling operation introduced by \cite{rudich_et_al:LIPIcs.CP.2022.35} can achieve.

\subsection{Memory Analysis}

\begin{figure}
    \centering
    \includegraphics[width=\textwidth]{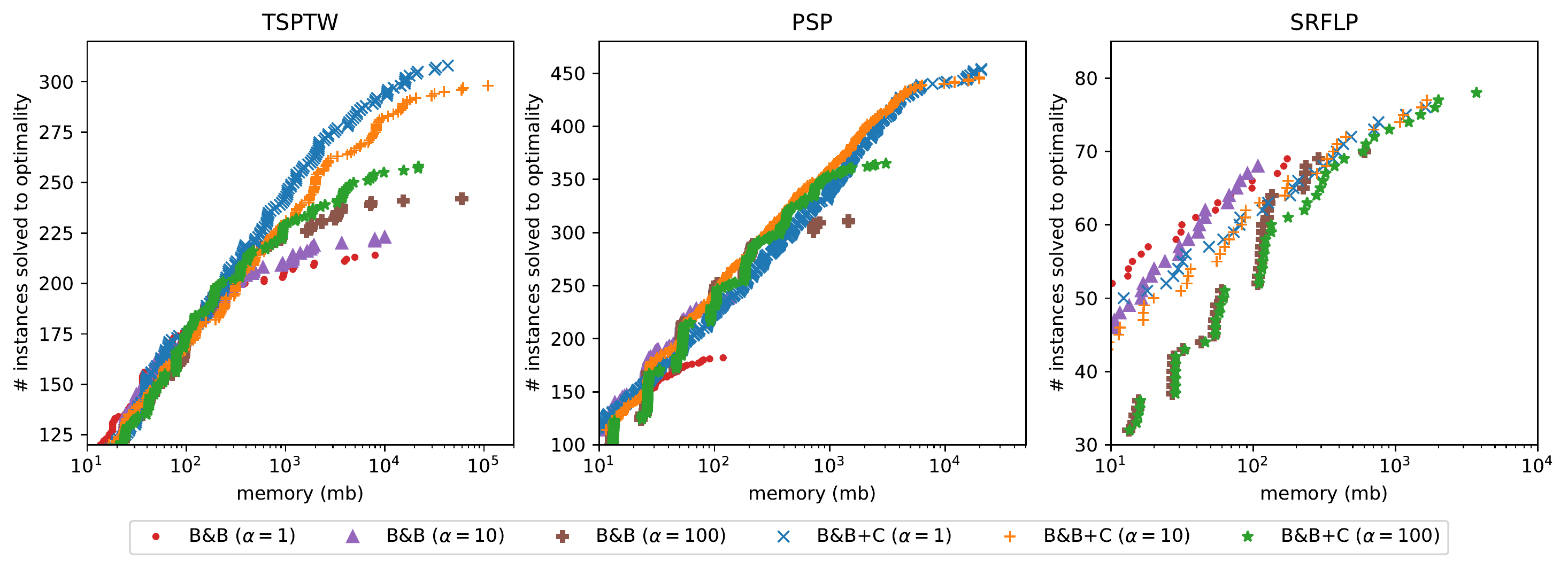}
    \caption{Number of instances solved by B\&B and B\&B+C for three different problems and for multiple maximum widths obtained with $\alpha \in \set{1,10,100}$, with respect to the peak amount of memory used.}
    \label{fig:results-memory}
\end{figure}

The performance improvements discussed in \Cref{results-speed} do not come completely for free, as the \textit{Cache} must store all the thresholds computed during the B\&B algorithm.
It is thus important to study the impact of this technique on the memory consumption of the algorithm.
For every instance solved by each algorithm, the peak amount of memory used during the execution was recorded.
\Cref{fig:results-memory} shows the number of instances solved using a given maximum amount of memory.
It appears that the memory consumption of B\&B and B\&B+C are of the same order of magnitude.
Actually, for the TSPTW and the PSP, as the instances get harder, B\&B+C even starts solving more instances with the same peak amount of memory.
Thus, even in terms of memory consumption, the cost of maintaining the \textit{Cache} seems to be compensated by its pruning effect which causes DDs to be sparser.

\section{Conclusion}
\label{section:conclusion}

In this paper, we first discussed how the DD-based B\&B algorithm tends to repeatedly explore overlapping parts of the search space.
Then, we introduced dominance and pruning thresholds with the intention of overcoming this limitation.
The former propagate dominance relations between partial solutions obtained within approximate DDs while the latter allow the approach to be combined and strengthened by the pruning performed by LocBs and RUBs.
Both types of thresholds are used in a single pruning mechanism that is able to discard many nodes associated with previously visited DP states during subsequent DD compilations.
Finally, we presented experimental results that clearly show the impact of the techniques introduced: B\&B with caching vastly outperforms the classical B\&B algorithm in terms of instances solved on the three discrete optimization problems studied in the paper, and compares well to state-of-the-art MIP models.
Furthermore, the experiments showed that the memory consumption induced by the \textit{Cache} was either acceptable or even overcompensated by its pruning effect.

In general, we expect that using the \textit{Cache} will be beneficial for solving problems whenever the theoretical search tree of the DP model comprises isomorphic subtrees, which are normally superimposed in the corresponding exact DD.
That is, we expect that the \textit{Cache} to be profitable whenever pure DP would be an efficient -- albeit impractical because of memory limitations -- method for solving the problem at hand.

DD-based B\&B is known to parallelize very well and it is one of its main advantages compared to other approaches.
The improvements described in this paper are based on a shared data structure that avoids successive DD compilations from being completely blind to previously done work.
Therefore, concurrent DD compilations are not fully independent anymore and require some synchronization to access and update the \textit{Cache}.
As future work, we thus aspire to optimize the pruning capacities of the \textit{Cache} while minimizing contention in the data structure.

%As future work, we aspire to combine the expansion thresholds described in this paper with the \textit{peel-and-bound} approach of \citep{rudich_et_al:LIPIcs.CP.2022.35}.
%We believe that these two improvements of the original DD-based B\&B algorithm can be combined and benefit from each other.
%Indeed, our caching mechanism would further reduce the overlap between successive approximate DD compilations for peel-and-bound.
%On the other hand, the peeling would allow to strengthen the upper bound of open nodes, which, as discussed in the experimental section, is currently lacking in our approach.

% Acknowledgments here
\ACKNOWLEDGMENT{%
We would like to thank Ryo Kuroiwa for providing us with his implementation of the MIP model for the TSPTW problem.
We also extend our thanks to the anonymous reviewers for their insightful comments and suggestions that helped improve the quality of the paper and clarify our contributions.
}% Leave this (end of acknowledgment)

% References here (outcomment the appropriate case) 

% CASE 1: BiBTeX used to constantly update the references 
%   (while the paper is being written).
\bibliographystyle{informs2014} % outcomment this and next line in Case 1
\bibliography{bibliography} % if more than one, comma separated

\begin{thebibliography}{55}
\providecommand{\natexlab}[1]{#1}
\providecommand{\url}[1]{\texttt{#1}}
\providecommand{\urlprefix}{URL }

\bibitem[{Akers(1978)}]{akers1978binary}
Akers SB (1978) Binary decision diagrams. \emph{IEEE Transactions on Computers}
  27(06):509--516.

\bibitem[{Amaral(2006)}]{amaral2006exact}
Amaral AR (2006) On the exact solution of a facility layout problem.
  \emph{European Journal of operational research} 173(2):508--518.

\bibitem[{Amaral(2008)}]{amaral2008exact}
Amaral AR (2008) An exact approach to the one-dimensional facility layout
  problem. \emph{Operations Research} 56(4):1026--1033.

\bibitem[{Amaral(2009)}]{amaral2009new}
Amaral AR (2009) A new lower bound for the single row facility layout problem.
  \emph{Discrete Applied Mathematics} 157(1):183--190.

\bibitem[{Andersen et~al.(2007)Andersen, Hadzic, Hooker, \protect\BIBand{}
  Tiedemann}]{andersen2007constraint}
Andersen HR, Hadzic T, Hooker JN, Tiedemann P (2007) A constraint store based
  on multivalued decision diagrams. \emph{International Conference on
  Principles and Practice of Constraint Programming}, 118--132 (Springer).

\bibitem[{Anjos et~al.(2005)Anjos, Kennings, \protect\BIBand{}
  Vannelli}]{anjos2005semidefinite}
Anjos MF, Kennings A, Vannelli A (2005) A semidefinite optimization approach
  for the single-row layout problem with unequal dimensions. \emph{Discrete
  Optimization} 2(2):113--122.

\bibitem[{Anjos \protect\BIBand{} Vannelli(2008)}]{anjos2008computing}
Anjos MF, Vannelli A (2008) Computing globally optimal solutions for single-row
  layout problems using semidefinite programming and cutting planes.
  \emph{INFORMS Journal on Computing} 20(4):611--617.

\bibitem[{Anjos \protect\BIBand{} Yen(2009)}]{anjos2009provably}
Anjos MF, Yen G (2009) Provably near-optimal solutions for very large
  single-row facility layout problems. \emph{Optimization Methods \& Software}
  24(4-5):805--817.

\bibitem[{Ascheuer(1996)}]{ascheuer1996hamiltonian}
Ascheuer N (1996) \emph{Hamiltonian path problems in the on-line optimization
  of flexible manufacturing systems}. Ph.D. thesis, University of Technology
  Berlin.

\bibitem[{Becker et~al.(2005)Becker, Behle, Eisenbrand, \protect\BIBand{}
  Wimmer}]{becker2005bdds}
Becker B, Behle M, Eisenbrand F, Wimmer R (2005) {BDDs} in a branch and cut
  framework. \emph{International Workshop on Experimental and Efficient
  Algorithms}, 452--463 (Springer).

\bibitem[{Bergman et~al.(2012)Bergman, Cire, van Hoeve, \protect\BIBand{}
  Hooker}]{bergman2012variable}
Bergman D, Cire AA, van Hoeve WJ, Hooker JN (2012) Variable ordering for the
  application of bdds to the maximum independent set problem.
  \emph{International conference on integration of artificial intelligence (AI)
  and operations research (OR) techniques in constraint programming}, 34--49
  (Springer).

\bibitem[{Bergman et~al.(2014{\natexlab{a}})Bergman, Cire, van Hoeve,
  \protect\BIBand{} Hooker}]{bergman2014optimization}
Bergman D, Cire AA, van Hoeve WJ, Hooker JN (2014{\natexlab{a}}) Optimization
  bounds from binary decision diagrams. \emph{INFORMS Journal on Computing}
  26(2):253--268.

\bibitem[{Bergman et~al.(2016)Bergman, Cire, van Hoeve, \protect\BIBand{}
  Hooker}]{bergman2016discrete}
Bergman D, Cire AA, van Hoeve WJ, Hooker JN (2016) Discrete optimization with
  decision diagrams. \emph{INFORMS Journal on Computing} 28(1):47--66.

\bibitem[{Bergman et~al.(2014{\natexlab{b}})Bergman, Cire, van Hoeve,
  \protect\BIBand{} Yunes}]{bergman2014bdd}
Bergman D, Cire AA, van Hoeve WJ, Yunes T (2014{\natexlab{b}}) {BDD}-based
  heuristics for binary optimization. \emph{Journal of Heuristics}
  20(2):211--234.

\bibitem[{Bryant(1986)}]{bryant1986graph}
Bryant RE (1986) Graph-based algorithms for boolean function manipulation.
  \emph{Computers, IEEE Transactions on} 100(8):677--691.

\bibitem[{Cappart et~al.(2022)Cappart, Bergman, Rousseau, Pr\'{e}mont-Schwarz,
  \protect\BIBand{} Parjadis}]{cappart2022improving}
Cappart Q, Bergman D, Rousseau LM, Pr\'{e}mont-Schwarz I, Parjadis A (2022)
  Improving variable orderings of approximate decision diagrams using
  reinforcement learning. \emph{INFORMS Journal on Computing} 34(5):2552--2570.

\bibitem[{Castro et~al.(2022)Castro, Cire, \protect\BIBand{}
  Beck}]{castro2022decision}
Castro MP, Cire AA, Beck JC (2022) Decision diagrams for discrete optimization:
  A survey of recent advances. \emph{INFORMS Journal on Computing}
  34(4):2271--2295.

\bibitem[{Cire \protect\BIBand{} van Hoeve(2013)}]{cire2013multivalued}
Cire AA, van Hoeve WJ (2013) Multivalued decision diagrams for sequencing
  problems. \emph{Operations Research} 61(6):1411--1428.

\bibitem[{Clarke et~al.(1994)Clarke, Grumberg, \protect\BIBand{}
  Long}]{clarke1994model}
Clarke EM, Grumberg O, Long DE (1994) Model checking and abstraction. \emph{ACM
  transactions on Programming Languages and Systems (TOPLAS)} 16(5):1512--1542.

\bibitem[{Copp\'{e} et~al.(2022)Copp\'{e}, Gillard, \protect\BIBand{}
  Schaus}]{coppe_et_al:LIPIcs.CP.2022.14}
Copp\'{e} V, Gillard X, Schaus P (2022) {Solving the Constrained Single-Row
  Facility Layout Problem with Decision Diagrams}. Solnon C, ed., \emph{28th
  International Conference on Principles and Practice of Constraint Programming
  (CP 2022)}, 14:1--14:18 (Schloss Dagstuhl -- Leibniz-Zentrum f{\"u}r
  Informatik).

\bibitem[{Copp{\'e} et~al.(2024)Copp{\'e}, Gillard, \protect\BIBand{}
  Schaus}]{DdoCaching}
Copp{\'e} V, Gillard X, Schaus P (2024) Decision diagram-based branch-and-bound
  with caching for dominance and suboptimality detection.
  \urlprefix\url{http://dx.doi.org/10.1287/ijoc.2022.0340.cd}, available for
  download at https://github.com/INFORMSJoC/2022.0340.

\bibitem[{Duff et~al.(1989)Duff, Grimes, \protect\BIBand{}
  Lewis}]{duff1989sparse}
Duff IS, Grimes RG, Lewis JG (1989) Sparse matrix test problems. \emph{ACM
  Transactions on Mathematical Software (TOMS)} 15(1):1--14.

\bibitem[{Dumas et~al.(1995)Dumas, Desrosiers, Gelinas, \protect\BIBand{}
  Solomon}]{dumas1995optimal}
Dumas Y, Desrosiers J, Gelinas E, Solomon MM (1995) An optimal algorithm for
  the traveling salesman problem with time windows. \emph{Operations research}
  43(2):367--371.

\bibitem[{Gendreau et~al.(1998)Gendreau, Hertz, Laporte, \protect\BIBand{}
  Stan}]{gendreau1998generalized}
Gendreau M, Hertz A, Laporte G, Stan M (1998) A generalized insertion heuristic
  for the traveling salesman problem with time windows. \emph{Operations
  Research} 46(3):330--335.

\bibitem[{Gillard et~al.(2021{\natexlab{a}})Gillard, Copp{\'e}, Schaus,
  \protect\BIBand{} Cire}]{gillard2021improving}
Gillard X, Copp{\'e} V, Schaus P, Cire AA (2021{\natexlab{a}}) Improving the
  filtering of branch-and-bound mdd solver. \emph{International Conference on
  Integration of Constraint Programming, Artificial Intelligence, and
  Operations Research}, 231--247 (Springer).

\bibitem[{Gillard \protect\BIBand{} Schaus(2022)}]{ijcai2022p659}
Gillard X, Schaus P (2022) Large neighborhood search with decision diagrams.
  De~Raedt L, ed., \emph{Proceedings of the Thirty-First International Joint
  Conference on Artificial Intelligence, {IJCAI-22}}, 4754--4760 (International
  Joint Conferences on Artificial Intelligence Organization).

\bibitem[{Gillard et~al.(2021{\natexlab{b}})Gillard, Schaus, \protect\BIBand{}
  Copp{\'e}}]{gillard2021ddo}
Gillard X, Schaus P, Copp{\'e} V (2021{\natexlab{b}}) {DDO}, a generic and
  efficient framework for mdd-based optimization. \emph{Proceedings of the
  Twenty-Ninth International Conference on International Joint Conferences on
  Artificial Intelligence}, 5243--5245.

\bibitem[{{Gurobi Optimization, LLC}(2022)}]{gurobi}
{Gurobi Optimization, LLC} (2022) {Gurobi Optimizer Reference Manual}.
  \urlprefix\url{https://www.gurobi.com}.

\bibitem[{Hachtel \protect\BIBand{} Somenzi(1997)}]{Hachtel1997}
Hachtel GD, Somenzi F (1997) A symbolic algorithms for maximum flow in 0-1
  networks. \emph{Formal Methods in System Design} 10(2):207--219.

\bibitem[{Had{\v{z}}i{\'{c}} \protect\BIBand{} Hooker(2006)}]{hadzic2006}
Had{\v{z}}i{\'{c}} T, Hooker JN (2006) Postoptimality analysis for integer
  programming using binary decision diagrams. Technical report, Carnegie Mellon
  University.

\bibitem[{Had{\v{z}}i{\'{c}} \protect\BIBand{} Hooker(2007)}]{hadzic2007}
Had{\v{z}}i{\'{c}} T, Hooker JN (2007) Cost-bounded binary decision diagrams
  for 0-1 programming. \emph{Integration of AI and OR Techniques in Constraint
  Programming for Combinatorial Optimization Problems}, 84--98 (Springer).

\bibitem[{Hart et~al.(1968)Hart, Nilsson, \protect\BIBand{}
  Raphael}]{hart1968formal}
Hart PE, Nilsson NJ, Raphael B (1968) A formal basis for the heuristic
  determination of minimum cost paths. \emph{IEEE transactions on Systems
  Science and Cybernetics} 4(2):100--107.

\bibitem[{Held \protect\BIBand{} Karp(1962)}]{held1962dynamic}
Held M, Karp RM (1962) A dynamic programming approach to sequencing problems.
  \emph{Journal of the Society for Industrial and Applied mathematics}
  10(1):196--210.

\bibitem[{Heragu \protect\BIBand{} Kusiak(1991)}]{heragu1991efficient}
Heragu SS, Kusiak A (1991) Efficient models for the facility layout problem.
  \emph{European Journal of Operational Research} 53(1):1--13.

\bibitem[{Hooker(2013)}]{hooker2013decision}
Hooker JN (2013) Decision diagrams and dynamic programming. \emph{International
  Conference on Integration of Constraint Programming, Artificial Intelligence,
  and Operations Research}, 94--110 (Springer).

\bibitem[{Hu(1995)}]{hu1995techniques}
Hu AJ (1995) \emph{Techniques for efficient formal verification using binary
  decision diagrams}. Ph.D. thesis, Stanford University, Department of Computer
  Science.

\bibitem[{Hungerl{\"a}nder \protect\BIBand{}
  Rendl(2013)}]{hungerlander2013computational}
Hungerl{\"a}nder P, Rendl F (2013) A computational study and survey of methods
  for the single-row facility layout problem. \emph{Computational Optimization
  and Applications} 55(1):1--20.

\bibitem[{Hungerl{\"a}nder \protect\BIBand{}
  Truden(2018)}]{hungerlander2018efficient}
Hungerl{\"a}nder P, Truden C (2018) Efficient and easy-to-implement
  mixed-integer linear programs for the traveling salesperson problem with time
  windows. \emph{Transportation research procedia} 30:157--166.

\bibitem[{Kam \protect\BIBand{} Brayton(1990)}]{kam1990multi}
Kam TYk, Brayton RK (1990) \emph{Multi-valued decision diagrams} (Electronics
  Research Laboratory, College of Engineering, University of California).

\bibitem[{Karahalios \protect\BIBand{} van
  Hoeve(2022)}]{karahalios2022variable}
Karahalios A, van Hoeve WJ (2022) Variable ordering for decision diagrams: A
  portfolio approach. \emph{Constraints} 27(1):116--133.

\bibitem[{{Lai} et~al.(1994){Lai}, {Pedram}, \protect\BIBand{}
  {Vrudhula}}]{lai1994evbdd}
{Lai} YT, {Pedram} M, {Vrudhula} SBK (1994) {EVBDD}-based algorithms for
  integer linear programming, spectral transformation, and function
  decomposition. \emph{IEEE Transactions on Computer-Aided Design of Integrated
  Circuits and Systems} 13(8):959--975.

\bibitem[{Langevin et~al.(1993)Langevin, Desrochers, Desrosiers, G{\'e}linas,
  \protect\BIBand{} Soumis}]{langevin1993two}
Langevin A, Desrochers M, Desrosiers J, G{\'e}linas S, Soumis F (1993) A
  two-commodity flow formulation for the traveling salesman and the makespan
  problems with time windows. \emph{Networks} 23(7):631--640.

\bibitem[{Lee(1959)}]{lee1959representation}
Lee CY (1959) Representation of switching circuits by binary-decision programs.
  \emph{The Bell System Technical Journal} 38(4):985--999.

\bibitem[{Minato(1995)}]{minato1995binary}
Minato Si (1995) \emph{Binary decision diagrams and applications for VLSI CAD},
  volume 342 (Springer Science \& Business Media).

\bibitem[{Nugent et~al.(1968)Nugent, Vollmann, \protect\BIBand{}
  Ruml}]{nugent1968experimental}
Nugent CE, Vollmann TE, Ruml J (1968) An experimental comparison of techniques
  for the assignment of facilities to locations. \emph{Operations research}
  16(1):150--173.

\bibitem[{Obata(1981)}]{obata1980quadratic}
Obata T (1981) The quadratic assignment problem: {Evaluation} of exact and
  heuristic algorithms. \emph{Transportation Research Part A: General}
  15(4):346.

\bibitem[{Ohlmann \protect\BIBand{} Thomas(2007)}]{ohlmann2007compressed}
Ohlmann JW, Thomas BW (2007) A compressed-annealing heuristic for the traveling
  salesman problem with time windows. \emph{INFORMS Journal on Computing}
  19(1):80--90.

\bibitem[{Pesant et~al.(1998)Pesant, Gendreau, Potvin, \protect\BIBand{}
  Rousseau}]{pesant1998exact}
Pesant G, Gendreau M, Potvin JY, Rousseau JM (1998) An exact constraint logic
  programming algorithm for the traveling salesman problem with time windows.
  \emph{Transportation Science} 32(1):12--29.

\bibitem[{Pochet \protect\BIBand{} Wolsey(2006)}]{pochet2006production}
Pochet Y, Wolsey LA (2006) \emph{Production planning by mixed integer
  programming}, volume 149 (Springer).

\bibitem[{Potvin \protect\BIBand{} Bengio(1996)}]{potvin1996vehicle}
Potvin JY, Bengio S (1996) The vehicle routing problem with time windows part
  {II}: genetic search. \emph{INFORMS Journal on Computing} 8(2):165--172.

\bibitem[{Rudich et~al.(2022)Rudich, Cappart, \protect\BIBand{}
  Rousseau}]{rudich_et_al:LIPIcs.CP.2022.35}
Rudich I, Cappart Q, Rousseau LM (2022) {Peel-And-Bound: Generating Stronger
  Relaxed Bounds with Multivalued Decision Diagrams}. Solnon C, ed., \emph{28th
  International Conference on Principles and Practice of Constraint Programming
  (CP 2022)}, 35:1--35:20 (Schloss Dagstuhl -- Leibniz-Zentrum f{\"u}r
  Informatik).

\bibitem[{Sarker(1989)}]{sarker1989amoebic}
Sarker BR (1989) \emph{The amoebic matrix and one-dimensional machine location
  problems} (Texas A\&M University).

\bibitem[{Simmons(1969)}]{simmons1969one}
Simmons DM (1969) One-dimensional space allocation: an ordering algorithm.
  \emph{Operations Research} 17(5):812--826.

\bibitem[{Wegener(2000)}]{wegener2000}
Wegener I (2000) Branching programs and binary decision diagrams: Theory and
  applications. \emph{Discrete Applied Mathematics} .

\bibitem[{Yu \protect\BIBand{} Sarker(2003)}]{yu2003directional}
Yu J, Sarker BR (2003) Directional decomposition heuristic for a linear
  machine-cell location problem. \emph{European Journal of Operational
  Research} 149(1):142--184.

\end{thebibliography}

% CASE 2: BiBTeX used to generate mypaper.bbl (to be further fine tuned)
%\input{mypaper.bbl} % outcomment this line in Case 2

\begin{APPENDIX}{Models Used in the Experimental Study}

The three discrete optimization problems for which we present models in this appendix are minimization problems.
The DD-based optimization framework remains applicable but the terminology must be adapted.
For instance, the optimal solution of a minimization problem is given by the \textit{shortest} path in the corresponding exact DD.
Moreover, restricted and relaxed DDs respectively provide upper and lower bounds.
Similarly, rough upper bounds are replaced with \textit{rough lower bounds} (RLBs).

\section{TSPTW}
\label{tsptw}

The TSPTW is a variant of the well-known Traveling Salesman Problem (TSP) where the cities are replaced by a set of customers $N=\set{\fromto{0}{n-1}}$ that must each be visited during a given time window $\mathcal{TW}_i = (e_i, l_i)$.
The first customer is a dummy customer that represents the depot where the salesman must begin and end its tour.
As in the classical TSP, we are given a symmetrical distance matrix $D$ that contains in each entry $(i,j)$ the distance $D_{ij}$ separating customers $i$ and $j$.
In addition to the time windows controlling the earliest and latest time when the salesman can visit the customers, the horizon $H$ limits the time at which the salesman must return to the depot.

\subsection{DP Model}
\label{tsptw:dp}

The DP model presented here extends the one introduced by \cite{held1962dynamic} for the TSP, which successively decides the next customer to visit and defines states with the set of customers that still must be visited along with the current position of the salesman.
For the TSPTW, this state representation is extended to a tuple $\vct{L, t, M, P}$, 
where $L$ (locations) and $t$ (time) respectively represent the set of locations where the salesman might be and the minimum time to reach one of these locations. 
The set $M$ (must) contains the visits that must still be completed, while $P$ (possible) are the visits that can possibly be done in addition to the ones in $M$. 
This set of possible visits $P$ is useful for tightening the relaxation of the relaxed MDD, as explained next.
$M$ and $P$ are disjoint sets of customers that must or might be visited in order to close the tour even in case of relaxed states.
We now describe the full DP model:
\begin{itemize}
    \item Control variables: $x_j \in N$ with $j \in N$ decides which customer is visited in $j$-th position.
    \item State space: $S = \set{\vct{L, t, M, P} \mid L, M, P \subseteq N, M \cap P = \emptyset, 0 \le t \le H}$.
    The root state is $\hat{r} = \vct{\set{0}, 0, N, \emptyset}$ and the terminal states are all states $\vct{\set{0}, t, \emptyset, P}$ with $0 \le t \le H$.
    \item Transition functions:
    \begin{equation}
    \label{tsptw:transition}
        t_j(s^j,x_j) = \left\{ \begin{array}{lcl}
                \vct{t_j^{L}(s^j,x_j), t_j^{t}(s^j,x_j), t_j^{M}(s^j,x_j), t_j^{P}(s^j,x_j)}, &&
                    \begin{array}{ll}
                        \text{if $x_j \in s^j.M$} \\
                        \text{and $s^j.t + \min_{i \in s^j.L} D_{ix_j} \le l_{x_j}$},
                    \end{array} \\
                \vct{t_j^{L}(s^j,x_j), t_j^{t}(s^j,x_j), t_j^{M}(s^j,x_j), t_j^{P}(s^j,x_j)}, &&
                    \begin{array}{ll}
                        \text{if $x_j \in s^j.P$} \\
                        \text{and $s^j.t + \min_{i \in s^j.P} D_{ix_j} \le l_{x_j}$}\\
                        \text{and $\rvert M \rvert < n - j$},
                    \end{array} \\
            \hat{0}, && \text{otherwise.}
            \end{array} \right.
    \end{equation}
    where
    $$
    \begin{array}{ll}
        t_j^{L}(s^j,x_j) &= \set{x_j} \\
        t_j^{t}(s^j,x_j) &= \max \set{ e_{x_j}, s^j.t + \min_{i \in s^j.P} D_{ix_j}} \\
        t_j^{M}(s^j,x_j) &= s^j.M \setminus \set{x_j} \\
        t_j^{P}(s^j,x_j) &= s^j.P \setminus \set{x_j}
    \end{array}
    $$
    In \Cref{tsptw:transition}, transitions are only allowed if they respect the time window constraint $l_{x_j}$.
    Since the salesman can be at multiple different positions in relaxed states, the minimum distance is used to compute the arrival time.
    The second condition in \Cref{tsptw:transition} ensures that no customers are selected from the possible set $P$ when there remains enough stops only for the customers in the must-set $M$.
    \item Transition value functions: $
        h_j(s^j,x_j) = \min_{i \in s^j.L} D_{ix_j}$.
    \item Root value: $v_r = 0$.
\end{itemize}

\subsection{Relaxation}
\label{tsptw:relax}

The merging operator is defined as follows:
\begin{equation}
    \oplus(\mathcal{M}) = \vct{\oplus_{L}(\mathcal{M}), \oplus_{t}(\mathcal{M}), \oplus_{M}(\mathcal{M}), \oplus_{P}(\mathcal{M})}
\end{equation}
where
$$
\begin{array}{ll}
    \oplus_{L}(\mathcal{M}) &= \bigcup_{s \in \mathcal{M}} s.L\\
    \oplus_{t}(\mathcal{M}) &= \min_{s \in \mathcal{M}} s.t \\
    \oplus_{M}(\mathcal{M}) &= \bigcap_{s \in \mathcal{M}} s.M \\
    \oplus_{P}(\mathcal{M}) &= (\bigcup_{s \in \mathcal{M}} s.M \cup s.P) \setminus (\bigcap_{s \in \mathcal{M}} s.M)
\end{array}
$$
These operators ensure that all transitions are preserved and that the transition values are not increased.
Note that the $M$ and $P$ sets are disjoint so that $P$ contains customers that must or might be visited in some states, but no customers that must be visited in all states.
The relaxed transition value operator is simply the identity function $\Gamma_\mathcal{M}(v,u) = v$.

\subsection{Rough Lower Bound}

The RLB shown in \Cref{tsptw:rlb} computes an estimate of the remaining distance to cover in order for the salesman to finish his tour and return to the depot.
For all customers $i$ in $M$, the distance of the cheapest edge incident to $i$ is added.
Then, $n - j - \lvert M \rvert$ customers from $P$ must be visited in order to complete the tour.
Since this is a lower bound computation, we create a vector $C$ containing the cheapest edge incident to each customer in $P$ and select the $n - j - \lvert M \rvert$ shortest ones.
We denote by $X_{<,i}$ the $i$-th smallest element from a vector $X$.
\begin{equation}
\label{tsptw:rlb}
    \lb{v}_{rlb}(s^j) = \sum_{i \in s^j.M} cheapest_i + \sum_{i = 1}^{n - j - \lvert M \rvert} C_{<,i}
\end{equation}
A check is also added to detect states where it is impossible to visit independently all of the customers from $M$ and at least $n-j-\lvert M \rvert$ customers from $P$, given the current time and the time windows of the remaining customers.
Plus, if the estimate provided by \Cref{tsptw:rlb} prevents the salesman from returning to the depot within the time constraint, a RLB of $-\infty$ is returned, effectively discarding the corresponding states.

\section{PSP}

The PSP is a single-machine planning problem where one has to find a production schedule that satisfies a set of demands at minimal cost.
Formally, there is a set of item types $I = \set{\fromto{0}{n-1}}$ that are associated with a stocking cost $S_i$.
When the machine switches the production from item type $i$ to  $j$, it incurs a changeover cost $C_{ij}$.
The planning spans for a given time horizon $H$.
For each time period $0 \le p < H$ of this horizon, $Q_p^i \in \set{0, 1}$ indicates whether an item of type $i$ must be delivered.
When an item $i$ is produced at time period $p_1$ and delivered at period $p_2$, stocking the item costs $S_i(p_2 - p_1)$.

To give a better understanding of the problem, we hereby recall the MIP model denoted PIG-A-1 in \citep{pochet2006production}.
Variables $x^i_p \in \set{0,1}$ decide whether an item of type $i$ is produced at time period $p$.
On the other hand, variables $y^i_p \in \set{0,1}$ decide whether the machine is ready to produce an item of type $i$ at time period $p$.
Indeed, the machine can be idle at certain periods.
Variables $q^i_p \in \mathbb{N}_0$ accumulate the quantity of items of type $i$ stored at period $p$.
Finally, variables $\chi^{i,j}_p \in \set{0,1}$ capture a changeover between item types $i$ and $j$ at period $p$.
\begin{align}
    \min & \sum_{i \in I} \sum_{p=0}^{H-1} S_i q^i_p + \sum_{i,j \in I} \sum_{p=0}^{H-1} C_{ij} \chi^{i,j}_p & \\
    & q^i_{p-1} + x^i_p = q^i_p + Q^i_p & \forall i \in I, 0 \le p < H \label{conservation-constraint} \\
    & q^i_{-1} = 0 & \forall i \in I \label{init-constraint} \\
    & x^i_p \le y^i_p & \forall i \in I, 0 \le p < H \label{setup-constraint} \\
    & \sum_{i \in I} y^i_p = 1 & \forall 0 \le p < H \label{mode-constraint} \\
    & \chi^{i,j}_p \ge y^i_{p-1} + y^i_p - 1 & \forall i,j \in I, 0 < p < H \label{changeover-constraint}
\end{align}
\Cref{conservation-constraint} models the stocking of the items and consumes them when needed, with the quantities initialized by \Cref{init-constraint}.
Then, \Cref{setup-constraint} ensures that the machine is in the correct mode to produce an item and \Cref{mode-constraint} allows only one mode for each time period.
Finally, \Cref{changeover-constraint} sets the correct value for the changeover variables.

\subsection{DP Model}

The DP model introduced in \citep{ijcai2022p659} is extended to allow relaxation as well as to cover instances where the machine can be idle at some time periods.
It proceeds by taking decisions starting from the end of the planning horizon and working backwards to avoid expanding any infeasible states.
For clarity, variables $x_j$ decide the type of item produced at period $j$.
This implies that the reverse variable ordering $\fromto{x_{H-1}}{x_0}$ is used.
Before pursuing the modeling of the PSP in the DD framework, let us define $P^i_r$ as the time period at which an item of type $i$ must be delivered when there remains $r$ of them to produce, i.e., $P^i_r = \min \set{0 \le q < H \mid \sum_{p=0}^q Q_p^i \ge r}$ for all $i \in N, 0 \le r \le \sum_{0 \le p < H} Q_p^i$.

States $\vct{i, R}$ where $i$ is the item type produced at the next time period and $R$ is a vector where $R_i$ gives the remaining number of demands to satisfy for item type $i$.
We pose $N' = N \cup \set{\bot}$ where $\bot$ is a dummy item type used to represent periods where the machine is idle.

\begin{itemize}
    \item Control variables: $x_j \in N'$ with $0 \le j < H$ decides which item type is produced at time period $j$.
    \item State space: $S = \set{s \mid s.i \in N', \forall i \in N, 0 \le s.R_i \le \sum_{0 \le p < H} Q_p^i}$.
    The root state is $\hat{r} = \vct{\bot, (\fromto{\sum_{0 \le p < H} Q_p^0}{\sum_{0 \le p < H} Q_p^{n-1}})}$ and the terminal states are of the form $\vct{i, (\fromto{0}{0})}$ with $i \in N'$.
    \item Transition functions:
    \begin{equation}
    \label{psp:transition}
        t_j(s^j,x_j) = \left\{ \begin{array}{lcl}
                \vct{t_j^{i}(s^j,x_j), t_j^{R}(s^j,x_j)}, && \text{if $x_j \ne \bot$ and $s^j.R_{x_j} > 0$ and $j \le P^{x_j}_{s^j.R_{x_j}}$}, \\
                \vct{t_j^{i}(s^j,x_j), t_j^{R}(s^j,x_j)}, && \text{if $x_j = \bot$ and $\sum_{i \in N} s^j.R_i < j + 1$}, \\
            \hat{0}, && \text{otherwise.}
            \end{array} \right.
    \end{equation}
    where
    $$
    \begin{array}{ll}
        t_j^{i}(s^j,x_j) &= \left\{\begin{array}{lcl}
            x_j, && \text{if $x_j \ne \bot$} \\
            s^j.i, && \text{otherwise.}
        \end{array}\right. \\
        t_j^{R}(s^j,x_j) &= \left\{\begin{array}{lcl}
            (s^j.R_0,\ldots,s^j.R_{x_j} - 1,\ldots,s^j.R_{n-1}), && \text{if $x_j \ne \bot$} \\
            s^j.R, && \text{otherwise.}
        \end{array}\right. 
    \end{array}
    $$
    The first condition in \Cref{psp:transition} states that in order to schedule a production of item type $x_j$ at state $s^j$, there must remain at least one demand of that item type to satisfy, and due at period $j$ or after.
    The second one ensures that idle periods can only be scheduled when the total remaining demand is smaller than the number of remaining time periods.
    \item Transition value functions:
    \begin{equation*}
        h_j(s^j,x_j) = \left\{\begin{array}{lcl}
            C_{x_js^j.i}, && \text{if $x_j \ne \bot$ and $s^j.i \ne \bot$} \\
            0, && \text{otherwise.}
        \end{array}\right\}
        +
        \left\{\begin{array}{lcl}
            S_{x_j} \cdot (j - P^{x_j}_{s^j.R_{x_j}}), && \text{if $x_j \ne \bot$} \\
            0, && \text{otherwise.}
        \end{array}\right\}
    \end{equation*}
    \item Root value: $v_r = 0$.
\end{itemize}

\subsection{Relaxation}

The merging operator is defined as follows:
\begin{equation}
    \oplus(\mathcal{M}) = \vct{\bot, (\fromto{\min_{s \in \mathcal{M}} s.R_0}{\min_{s \in \mathcal{M}} s.R_{n-1}})}.
\end{equation}
As merged states might disagree on the item type produced before, it is reset to $\bot$.
For each item type, the minimum remaining number of items to produce is computed, meaning that all demands satisfied by a least one state are considered satisfied in the merged state.
As for the TSPTW, the relaxed transition value operator is the identity function $\Gamma_\mathcal{M}(v,u) = v$.

\subsection{Rough Lower Bound}

When the changeover costs are ignored, the PSP falls under the Wagner-Whitin conditions \citep{pochet2006production} that allow computing the optimal stocking cost.
Conversely, if the stocking costs and the delivery constraints are omitted, the PSP can be reduced to the TSP.
Therefore, a valid lower bound on the total changeover cost to produce a remaining set of items is to take the total weight of a Minimum Spanning Tree computed on the graph of changeover costs limited to item types that still need to be produced.
The optimal weight for all these spanning trees can be precomputed because the number of items is usually small.
As there is no overlap between the two lower bounds described, the RLB for the PSP can sum their individual contributions to obtain a stronger lower bound.

\section{SRFLP}

The SRFLP is an ordering problem that aims to place a set of departments $N=\set{\fromto{0}{n-1}}$ on a line.
Each department $i \in N$ has a positive length $L_i$, and the connection between each pair of departments $i,j \in N$ is described by a positive traffic intensity $C_{ij}$.
The goal is to find a bijection $\pi : N \rightarrow N$ that maps each department to a position on the line, while minimizing the total distance covered in the facility:
\begin{equation}
\label{srflp:objective}
    SRFLP(\pi) = \sum_{k=0}^{n-1} L_k \sum_{\substack{i=0\\\pi(i) < \pi(k)}}^{n-1} \sum_{\substack{j=0\\\pi(k) < \pi(j)}}^{n-1} C_{ij} + \underbrace{\sum_{i=0}^{n-1} \sum_{\substack{j=i+1}}^{n-1} C_{ij}\frac{L_i+L_j}{2}}_{K}.
\end{equation}
The second term of \Cref{srflp:objective} is a constant that does not depend on the ordering $\pi$, which is usually denoted $K$.
The modeling used in this paper is similar to the one presented in \citep{coppe_et_al:LIPIcs.CP.2022.14} except that merging is included.
The following definitions were thus adapted to cover states where each department must, might or might not be added to the ordering, similarly to the TSPTW modeling explained in \Cref{tsptw}.

\subsection{DP Model}

The DP model works as follows: departments are added one by one on the line from left to right.
The states are defined by tuples $\vct{M, P, C}$ where $M$ (must) and $P$ (possible) are the sets of departments that respectively must and might be placed on the line.
$C$ (cuts) is a vector that accumulates the \textit{cut value} of each department.
The cut value of a department is the total traffic intensity incident to this department, coming from placed departments.
Formally, the DP model is defined as follows:

\begin{itemize}
    \item Control variables: $x_j \in N$ with $j \in N$ decides which department is placed at position $j$ on the line.
    \item State space: $S = \set{\vct{M, P, C} \mid M,P \subseteq N, M \cap P = \emptyset}$.
    The root state is $\hat{r} = \vct{N, \emptyset, (\fromto{0}{0})}$ and terminal states $\vct{M, P, C}$ must only verify $M = \emptyset$.
    \item Transition functions:
    \begin{equation}
    \label{srflp:transition}
        t_j(s^j,x_j) = \left\{ \begin{array}{lcl}
                \vct{t_j^{M}(s^j,x_j), t_j^{P}(s^j,x_j), t_j^{C}(s^j,x_j)}, && \text{if $x_j \in s^j.M$}, \\
                \vct{t_j^{M}(s^j,x_j), t_j^{P}(s^j,x_j), t_j^{C}(s^j,x_j)}, && \text{if $x_j \in s^j.P$ and $\lvert M \rvert < n - j$}, \\
            \hat{0}, && \text{otherwise.}
            \end{array} \right.
    \end{equation}
    where
    $$
    \begin{array}{lll}
        t_j^{M}(s^j,x_j) &= s^j.M \setminus \set{x_j} & \\
        t_j^{P}(s^j,x_j) &= s^j.P \setminus \set{x_j} & \\
        t_j^{C}(s^j,x_j) &= (\fromto{C_0}{C_{n-1}}) &
        \text{with } C_i = \left\{ \begin{array}{lcl}
            s^j.C_i + C_{x_ji}, && \text{if $x_j \in (s^j.M \cup s^j.P) \setminus \set{x_j} $}, \\
            0, && \text{otherwise.}
        \end{array} \right.
    \end{array}
    $$
    The second condition in \Cref{srflp:transition} ensures that no departments can be added from $s^j.P$ when there remains space only for the departments in $s^j.M$.
    \item Transition value functions: the contribution of place department $x_j$ at position $j$ is its length $L_{x_j}$ multiplied by the sum of the cut values of departments that will be located to its right.
    These departments are contained in $s^j.M$ and in case the state is relaxed, in $s^j.P$.
    However, there might be more departments in $s^j.M \cup s^j.P$ than needed to complete the ordering.
    Selecting the $n-j-\lvert s^j.M \rvert$ smallest cut values among departments of $s^j.P$ will result in a lower bound on the value obtained by placing any subset of departments in $s^j.P$.
    For a transition from state $s^j$ with decision $x_j$, we also define $C^{P,x_j} = \lst{s^j.C_i \mid i \in s^j.P \setminus \set{x_j}}$ as the vector of cut values of departments in $s^j.P \setminus \set{x_j}$.
    Then, the transition value functions are:
    \begin{equation*}
        h_j(s^j,x_j) = L_{x_j} \left( \sum_{i \in s^j.M \setminus \set{x_j}} s^j.C_i + \sum_{i = 1}^{n-j-\lvert s^j.M \rvert} C^{P,x_j}_{<,i} \right).
    \end{equation*}
    \item Root value: $v_r = K$ (see \Cref{srflp:objective}).
\end{itemize}

\subsection{Relaxation}

The merging operator is defined as follows:
\begin{equation}
    \oplus(\mathcal{M}) = \vct{\oplus_{M}(\mathcal{M}), \oplus_{P}(\mathcal{M}), \oplus_{C}(\mathcal{M})}
\end{equation}
where
$$
\begin{array}{lll}
    \oplus_{M}(\mathcal{M}) &= \bigcap_{s \in \mathcal{M}} s.M & \\
    \oplus_{P}(\mathcal{M}) &= (\bigcup_{s \in \mathcal{M}} s.M \cup s.P) \setminus (\bigcap_{s \in \mathcal{M}} s.M) & \\
    \oplus_{C}(\mathcal{M}) &= (\fromto{C_0}{C_{n-1}}) &
    \text{with } C_i = \min_{s \in \mathcal{M}} \left\{ \begin{array}{lcl}
        s.C_i, && \text{if $i \in s.M \cup s.P$}, \\
        \infty, && \text{otherwise.}
    \end{array} \right\}
\end{array}
$$
The $M$ and $P$ sets are aggregated just as in the TSPTW relaxation of \Cref{tsptw:relax}.
Concerning the cut values, $\oplus_{C}$ simply keeps for each department the minimum cut value among the states that still must or might place the department considered.
Once again, the relaxed transition value operator is the identity function $\Gamma_\mathcal{M}(v,u) = v$.

\subsection{Rough Lower Bound}
\label{srflp:rlb}

The RLB used for the SRFLP is divided into separate bounds: the first computes the optimal ordering of the remaining departments with respect to their cut values only, while the second estimates the arrangement cost of the remaining departments, ignoring the departments that have already been placed.
The \textit{first-generation bound} introduced by \cite{simmons1969one} allows computing the ordering that minimizes the contribution of cut values with respect to departments placed on the left.
In that case, the optimal arrangement is given by ordering the departments by decreasing cut-to-length ratios.
The only obstacle to using this lower bound is that for a state $s^j$ all departments from $s^j.M$ will be included in the arrangement but only $n^{P} = n - j - n^{M}$ from $s^j.P$, with $n^{M} = \lvert s^j.M \rvert$.
As the optimal subset of departments to select from $s^j.P$ is unknown, we create $n^{P}$ artificial departments with the $n^{P}$ shortest lengths and smallest cut values among the departments in $s^j.P$.
A vector of ratios that contain tuples $\vct{r, c, l}$ is created for departments in $s^j.M$ and $s^j.P$, where $r = \frac{c}{l}$, $c$ is a cut value and $l$ is a length.
As stated by \Cref{ratio-artificial}, the $k$-th shortest length is combined with the $(n^{P} - k + 1)$-th smallest cut value as to minimize the lower bound formula given by \Cref{lbcut}.
\begin{align}
    L^{P} &= (L_k \mid k \in s^j.P) \\
    C^{P} &= (L_k \mid k \in s^j.P) \\
    R^{P} &= \lst{\vct{\frac{C^{P}_{<,n^{P} - k + 1}}{L^{P}_{<,k}}, C^{P}_{<,n^{P} - k + 1}, L^{P}_{<,k}} \mid k=\fromto{1}{n^{P}}} \label{ratio-artificial} \\
    R^{M} &= \lst{\vct{\frac{s^j.C_k}{L_k},s^j.C_k,L_k} \mid k \in s^j.M}\\
    R &= concat(R^{M}, R^{P})
\end{align}
The concatenation of both vectors of ratios is then used to compute the lower bound as in \citep{simmons1969one}.
\begin{equation}
\label{lbcut}
    LB_{cut}(s^j) = \sum_{k = 1}^{n-j} R_{>,k}.c \sum_{l = 1}^{k-1} R_{>,l}.l
\end{equation}

The second part of the RLB is a lower bound on the arrangement cost of $n^{M}+n^{P}$ departments from $s^j.M$ and $s^j.P$.
First, a subset of traffic intensities is selected: all those connecting two departments in $s^j.M$ and then the $n^{M}n^{P}$ lowest of those connecting one department from $s^j.M$ to one in $s^j.P$ and the $\frac{n^{P} (n^{P} - 1)}{2}$ lowest of those connecting two departments in $s^j.P$.
Second, the $n^{P}$ shortest department lengths from $s^j.P$ are combined with those from $s^j.M$.
\begin{align}
    T^{M,M} &= \lst{C_{kl} \mid k,l \in s^j.M, k < l} \\
    T^{M,P} &= \lst{C_{kl} \mid k \in s^j.M, l \in s^j.P} \\
    T^{P,P} &= \lst{C_{kl} \mid k,l \in s^j.P, k < l} \\
    T &= concat\left(
        \begin{array}{l}
            T^{M,L},\\ T^{M,P}_<[1 \ldots n^{M}n^{P}],\\ T^{P,P}_<[1 \ldots \frac{n^{P} (n^{P} - 1)}{2}]
        \end{array}
    \right) \\
    L^{M} &=  \lst{L_k \mid k \in s^j.M} \\
    L &= concat(L^{M}, L^{P}_<[1 \ldots n^{P}])
\end{align}
Given the traffic and length vectors $T$ and $L$, a lower bound on the arrangement cost can then be computed by multiplying the traffic intensities by an optimistic distance.
In \Cref{lbedge}, the lowest traffic intensity is multiplied by the $n-j-2$ shortest department lengths.
Then, the next two lowest traffic intensities are multiplied by the $n-j-3$ shortest department lengths, and so on.
With $\Delta_k = \frac{k(k+1)}{2}$ the $k$-th triangular number, we write:
\begin{equation}
\label{lbedge}
    LB_{edge}(s^j) = \sum_{k=1}^{n-j-1} \sum_{l=\Delta_{k-1}+1}^{\Delta_k} T_{<,l} \sum_{m=1}^{n-j-1-k} L_{<,m}.
\end{equation}
In the end, the RLB for the SRFLP is given by: $\lb{v}_{rlb}(s^j) = LB_{cut}(s^j) + LB_{edge}(s^j)$.

\end{APPENDIX}

\end{document}